\documentclass[epj]{svjour}
\usepackage{graphicx}
\usepackage{graphics}
\usepackage{amssymb}
\usepackage{siunitx}
\usepackage{amsmath}
\usepackage{multirow}
\usepackage{lineno}
\usepackage{float}
\usepackage{booktabs}
\usepackage{hyperref}
\usepackage{bm}
\usepackage{subfigure}

\begin{document}
\title{Determine Energy Nonlinearity and Resolution of $e^{\pm}$ and $\gamma$ in Liquid Scintillator Detectors by A Universal Energy Response Model}
\author{Miao Yu\inst{1} \and 
Liangjian Wen\inst{2}
\thanks{email: wenlj@ihep.ac.cn (corresponding author)} \and
Xiang Zhou\inst{1} \and
Wuming Luo\inst{2}
} 

\institute{School of Physics and Technology, Wuhan University, Wuhan 430072, China \and
Institute of High Energy Physics, Chinese Academy of Sciences, Beijing 100049, China}


\date{Received: date / Revised version: date}

\abstract{
Energy nonlinearity and resolution in liquid scintillator (LS) detectors are correlated and particle-dependent. A unified energy response model for liquid scintillator detectors has been presented in details. This model has advanced a data-driven approach to calibrate the particle-dependent energy response, using both the monoenergetic $\gamma$-ray sources and the continuous $\beta$ spectra of $^\mathrm{12}\mathrm{B}$ and Michel $e^-$ induced by cosmic muons. Monte Carlo studies have demonstrated the effectiveness and robustness of the proposed model, in particular, the positron energy resolution can be extracted in the absence of positron sources. This work will provide a feasible approach of simultaneous calibration of energy nonlinearity and resolution for the running and future LS detectors.
%
} 

\titlerunning{ } 
\authorrunning{ }
\maketitle
\section{Introduction}\label{Sec:intro}
Liquid scintillator (LS) detectors, composed of LS and photosensors, such as photomultiplier tubes (PMTs), have been widely used in neutrino experiments since the first detection of reactor antineutrinos~\cite{Reines:1953pu}. Among the reactor neutrino experiments, KamLAND has measured the $\theta_{12}$ driven oscillation~\cite{KamLAND:2008dgz}, and $\theta_{13}$ was measured at Daya Bay~\cite{DayaBay:2012fng}, RENO~\cite{RENO:2012mkc} and Double Chooz~\cite{DoubleChooz:2011ymz}. Besides, Borexino has made remarkable contributions in solar neutrino measurements~\cite{Borexino:2017rsf,BOREXINO:2020aww}. LS detectors also play important roles in the searches for neutrinoless double-beta ($0\nu\beta\beta$) decays, such as KamLAND-Zen~\cite{KamLAND-Zen:2012mmx} and SNO+~\cite{SNO:2021xpa}.In the future, JUNO will build the largest LS detector to determine neutrino mass ordering (NMO) by precisely measuring the fine oscillation patterns with reactor neutrinos, which requires an unprecedented energy resolution of $3\%$ at 1 MeV of visible energy~\cite{JUNO:2015zny,Zhan:2008id,Zhan:2009rs,JUNO:2021vlw}.

For LS detectors used in reactor neutrino experiments, $e^+$ from the inverse beta decay (IBD) interaction is the target signal, however, $\gamma$s instead of $\beta$s are the most common calibration sources. Charged particles $e^\pm$ lose energy mostly by ionizing and exciting the solvent molecules, while $\gamma$ generates $e^\pm$ firstly~\cite{ParticleDataGroup:2020ssz}. The deexcited molecules release scintillation
photons. Charged particles moving faster than light in the medium produce Cherenkov photons. Photons propagate in the detector and undergo series of optical processes until they hit PMTs and get converted into photoelectrons (PEs). The mapping between the number of photoelectrons ($N_\mathrm{pe}$) and the particle deposited energy reflects the energy response of the LS detector, which is crucial to the spectral analysis of neutrino oscillations. 

Due to different energy deposition processes, energy response in LS is particle-dependent. The nonlinearities in energy releases of $e^\pm$ and $\gamma$ have been thoroughly studied~\cite{DayaBay:2019fje,JUNO:2020xtj}. For LS detectors with relatively poor energy resolution, the reported energy resolution is usually calibrated with mono-energetic $\gamma$ sources without considering particle types, e.g., see ref.~\cite{DayaBay:2016ggj}. For LS detectors with high energy resolution, they will have the potential to distinguish the energy resolution curves of different particle types, and the spectral analysis for physics topics like NMO, solar neutrino and $0\nu\beta\beta$ decays, etc, will benefit from more comprehensive knowledge of energy resolution. Thus, it is necessary to construct a unified energy resolution model. 

The $N_\mathrm{pe}$ of mono-energetic charged particles like $e^\pm$ is determined by the underlying energy deposition and optical processes in LS. $\gamma$s can be described by $e^\pm$ because they deposit energies by generating primary $e^\pm$ firstly. However, each $\gamma$ has a different deposition mode, among which the multiplicities and energies of primary $e^\pm$ are different. Therefore, the $N_\mathrm{pe}$ distribution of mono-energetic $\gamma$s needs to be derived from the collection of all deposition modes instead of a one-dimensional probability density function (PDF) of the kinetic energy of the primary $e^\pm$ (e.g., see figure 7 in ref.~\cite{DayaBay:2019fje}). All in all, the nonlinearity and resolution are intrinsically correlated.

In this paper, we developed a unified energy model to describe particle-dependent energy response for LS detectors. Following a similar strategy as ref.~\cite{DayaBay:2019fje}, we construct the energy response model for $e^-$ firstly, then derive the $\gamma$ and $e^+$ models from it and calibrate the model with $\gamma$ sources and continuous $\beta$ spectra. By taking into account the dispersions in $\gamma$ energy deposition modes mentioned above, the model is capable to describe nonlinearity and resolution simultaneously. The structure of this paper is as follows: in section~\ref{Sec:MCSim} details of the Geant4-based~\cite{GEANT4:2002zbu} Monte Carlo simulation are presented. In section~\ref{Sec:connections}, we elaborate on the connections among particles in LS. The methods of model construction are described in section~\ref{Sec:ModelConst}. Then in section~\ref{Sec:FitResults} the calibration procedures and performances of the model are presented. Furthermore, we discuss the implications of calibration inputs and various energy resolution scenarios in section~\ref{Sec:Dis}. Finally we give some further remarks and summarize our main conclusions in section~\ref{Sec:Summary}.

\section{Monte Carlo configurations} \label{Sec:MCSim}
A set of Geant4-based (version 4.10.p02) Monte Carlo simulation software is developed. For simplicity, a liquid scintillator sphere is implemented as the target which is surrounded by photosensors. PMT geometry and its response are not accounted in this work in order to exclude the corresponding contribution to the LS energy model.

The Geant4 physics packages and custom codes jointly describe the physics processes in the detector. The low energy electromagnetic processes are described by the default Livermore model. A custom scintillation process covers the quenching effect, photon emission, absorption and re-emission. Other optical processes including Cherenkov process, Rayleigh scattering and boundary interactions are depicted with the official Geant4 codes. Besides, the optical parameters of the liquid scintillator are key to the optical propagation. Wavelength-dependent refractive indices are derived from measurement values and the dispersion relation~\cite{Zhou:2015gwa}, and the attenuation length and Rayleigh scattering length are taken from Ref~\cite{Gao:2013pua,Zhou:2015gwa,Wurm:2010ad}. Scintillation properties are taken from the measurements including emission spectrum~\cite{Zhang:2020mqz}, fluorescence quantum yield~\cite{Xiao:2010ternary,Ding:2015sys,Buck:2015jxa} and time profile~\cite{Li:2011time,OKeeffe:2011dex}. Flexible studies with varied Monte Carlo inputs are feasible using our simulation software.

By tuning the detection efficiency of sensitive detectors, the photon statistics then the energy resolution vary among different simulation samples. Series of different detector configurations have been simulated under this framework by covering the range of energy resolution of several typical large-scale liquid scintillator detectors, for example JUNO ($\sim 3.0\%$)~\cite{JUNO:2015zny}, Borexino ($\sim 6.0\%$)~\cite{BOREXINO:2020aww}, KamLAND ($\sim 6.5\%$)~\cite{KamLAND:2008dgz} and Daya Bay ($\sim 8.4\%$)~\cite{DayaBay:2012fng}. A baseline light yield is set as $1400/\,\mathrm{MeV}$ which corresponds to the energy resolution of positrons with zero kinetic energy around $2.9\%$ in the current simulation framework. The other three configurations with light yields scaled by $1/2$, $1/4$ and $1/8$ have the corresponding energy resolutions $3.8\%$, $5.5\%$ and $7.8\%$. It is worth mentioning that the real corresponding relation between the light yield and the energy resolution depends on the specific experimental details, the values mentioned above are obtained under the simulation framework in this study.

\section{Connections among \texorpdfstring{$\boldsymbol{e^\pm}$}{e+-} and \texorpdfstring{$\boldsymbol{\gamma}$}{gamma} in liquid scintillator}
\label{Sec:connections}
For generality and simplicity, all particles are assumed at the detector center and true $N_\mathrm{pe}$ detected by PMTs is used, so that the energy non-uniformity is not considered in this model. The detector-dependent instrumental effects require specific studies for each experiment and can be added to this model.

To convert $N_\mathrm{pe}$ into the energy dimension, a general definition of ``visible energy'' is introduced as,
\begin{eqnarray}
    E_\mathrm{vis} & \equiv & {N_\mathrm{pe}} / Y, \label{Eq:ParEvis}
\end{eqnarray}
where $Y$ is the photoelectrons yield per ${\rm MeV}$ and is calibrated by the neutron capture on hydrogen using neutron calibration sources. The energy-dependent nonlinearity ($f$) and resolution ($R$) for the mono-energetic particles are defined as,
\begin{eqnarray}
    f(E) &\equiv & \frac{\overline{E_\mathrm{vis}}}{E}, \label{Eq:ParNonl} \\
    R(E_\mathrm{vis}) & \equiv & 
    \frac{\sigma}{\overline{E_\mathrm{vis}}},
    \label{Eq:ParResol}
\end{eqnarray}
where $\overline{E_\mathrm{vis}}$ and $\sigma$ are the expected value and standard deviation of $E_\mathrm{vis}$, respectively. 

Similar to ref.~\cite{DayaBay:2019fje}, the $e^-$ is regarded as a basic particle, which has energy response described by Eq.~\eqref{Eq:ParNonl} and Eq.~\eqref{Eq:ParResol} directly. In our model, we consider the $e^+$ deposits its kinetic energy $T$ equivalently as $e^-$ approximately, followed by two identical $0.511\,\mathrm{MeV}$ annihilation $\gamma$s. Therefore, the energy response of $e^+$ can be predicted with that of $e^-$ directly,
\begin{eqnarray}
f^{e^+} & \equiv & \frac{\overline{E^{e^+}_\mathrm{vis}}}{E^{e^+}} = \frac{\overline{E^{e^-}_\mathrm{vis}}\left(T\right)+2\cdot \overline{E^{\gamma}_\mathrm{vis}}\left(m_e\right) }{T+2m_e},\label{Eq:e+NonlModel}\\
R^{e^+} & \equiv & \frac{\sigma^{e^+}}{\overline{E^{e^+}_\mathrm{vis}}} = \frac{\sqrt{\left[\sigma^{e^-}\left(T\right)\right]^2+ 2\left[\sigma^{\gamma}(m_e)\right]^2 }}{\overline{E^{e^-}_\mathrm{vis}}\left(T\right) + 2\cdot \overline{E^{\gamma}_\mathrm{vis}}\left(m_e\right) },
\label{Eq:e+ResModel}
\end{eqnarray}
where the kinetic energy and annihilation energies are considered as two independent parts. One way to describe the two $0.511\,\mathrm{MeV}$ annihilation $\gamma$s is employing the $^{\mathrm{68}}$Ge source. The $^{\mathrm{68}}$Ge source is a positron source while the kinetic energy is mostly deposited in the enclosure, and the two annihilation $\gamma$s drift out and deposit energy in LS. The explicit formulae can be found in section~\ref{Sec:ModelConst}.

There are two cases that might introduce systematic bias, annihilation in flight and formation of positronium. Annihilation in flight will generate two $\gamma$s with higher energies, which accounts for around $1\%$ per MeV from the simulation. If positronium is formed, there are two spin states: singlet (p-Ps) and triplet (o-Ps). Around $2\%$ triplets decay to three $\gamma$s with the total energy of $1.022\,\mathrm{MeV}$, while two $0.511\,\mathrm{MeV}$ $\gamma$s are released in the other cases. To include these effects into consideration, one way is to parameterize the energy response of $\gamma$. If one has the full knowledge of $\gamma$ energy distributions from above two effects, the energy response of $e^+$ can be corrected furthermore. In the simplified consideration of Eq.~\eqref{Eq:e+NonlModel}-\eqref{Eq:e+ResModel}, these two effects are ignored due to relatively small contributions.

The $\gamma$s deposit energy mainly by generating primary $e^\pm$ through three interactions: photoelectric effect, Compton scattering, and pair conversion. The multiplicities and energies of the primary $e^\pm$ from mono-energetic $\gamma$s are different event by event (named ``deposition mode'') resulting from the competition of those three interactions. According to the energy-dependent response functions in Eq.~\eqref{Eq:ParNonl} and Eq.~\eqref{Eq:ParResol}, each collection of primary $e^\pm$ in a deposition mode corresponds to a specific visible energy distribution:
\begin{eqnarray}
\left[E^\gamma_\mathrm{vis}\right]_j  =  \sum_{l^j}\left[E^{e^-}_\mathrm{vis}\right]_{l^j}+\sum_{n^j}\left[E^{e^+}_\mathrm{vis}\right]_{n^j}, \label{Eq:jthGamEvis}
\end{eqnarray}
where $j$ specifies the given deposition mode, while $l^j$ and $n^j$ enumerate all the primary $e^-$ and $e^+$ in the $j$-th deposition mode, respectively. The positron term in Eq.~\eqref{Eq:jthGamEvis} can be also described by the $e^-$ response according to Eq.~\eqref{Eq:e+NonlModel}-\eqref{Eq:e+ResModel}. The expected value of $\left[ E_\mathrm{vis}^\gamma \right]_j$ and its variance are calculated assuming that primary $e^\pm$ are independent:
\begin{align}
\overline{\left[E^\gamma_\mathrm{vis}\right]_j} & =  \sum_{l^j}\overline{\left[E^{e^-}_\mathrm{vis}\right]_{l^j}}+\sum_{n^j}\overline{\left[E^{e^+}_\mathrm{vis}\right]_{n^j}}, \label{Eq:jthGamEvisMean} \\
\left[\sigma^\gamma\right]^2_{j} & =  \sum_{l^j}\left[\sigma^{e^-}\right]^2_{l^j}+\sum_{n^j}\left[\sigma^{e^+}\right]^2_{n^j}. \label{Eq:jthGamEvisSigma} 
\end{align}

The total visible energy distribution is supposed to be a mixture distribution for all the deposition modes. In previous nonlinearity studies, only the expected value $\overline{E_\mathrm{vis}^\gamma}$ is considered thus a one-dimensional PDF averaging all deposition modes out is sufficient for calculation~\cite{DayaBay:2019fje}. However, to involve the resolution into consideration, each deposition mode requires to be calculated as an independent distribution so that the fluctuations among different modes are not washed off. Consequently, the nonlinearity and resolution for $\gamma$s can be calculated with the knowledge of all the possible deposition modes:

\begin{align}
f^\gamma & =  \frac{\overline{E_\mathrm{vis}^\gamma}}{E^\gamma} = \frac{\sum_j w_j \overline{\left[E^\gamma_\mathrm{vis}\right]_j}}{E^\gamma}, \label{Eq:gamNonl} \\
R^\gamma & =  \frac{\sigma^\gamma}{\overline{E_\mathrm{vis}^\gamma}}  \nonumber \\
& = \frac{\sqrt{\sum_jw_j\left[\overline{\left[E_\mathrm{vis}^\gamma\right]_j}-\overline{E_\mathrm{vis}^\gamma}\right]^2+ \sum_jw_j \left[\sigma^\gamma\right]^2_{j}}}{\overline{E_\mathrm{vis}^\gamma}}\label{Eq:gamRes} \\
& =  \frac{\sqrt{\sigma_\mathrm{nonl}^2 + \sigma_\mathrm{ave}^2}}{\overline{E_\mathrm{vis}^\gamma}}, \label{Eq:gamResDecomp}
\end{align}

where $w_j$ is the weight of the $j$-th deposition mode. The two terms in the numerator of Eq.~\eqref{Eq:gamRes} are redefined as $\sigma_\mathrm{nonl}^2$ and $\sigma_\mathrm{ave}^2$ in Eq.~\eqref{Eq:gamResDecomp}, respectively. The second term $\sigma_\mathrm{ave}^2$ is the weighted sum of variance for all primary $e^\pm$. The first term $\sigma_\mathrm{nonl}^2$ is the weighted sum of the square of mean value deviation which reflects the dispersion of all deposition modes. Apparently, this term requires an individual calculation of each event rather than the one-dimensional PDF, because 
different collections of the primary $e^\pm$ generated by the same $E^\gamma$ will transfer into different visible energy due to nonlinearity. The $\sigma_\mathrm{nonl}^2$ term will disappear if the energy response is linear in LS detectors. Therefore, the coupling of nonlinearity and resolution of $e^-$ is embedded in the $\gamma$ resolution.

\section{Model construction} \label{Sec:ModelConst}
According to discussions in section~\ref{Sec:connections}, the energy response for both $e^+$ and $\gamma$ can be deduced from that of $e^-$. Therefore, it is crucial to construct nonlinearity and resolution models for $e^-$ firstly. Details are described in the following.

\subsection{Energy response model of \texorpdfstring{$\boldsymbol{e^-}$}{e-}} \label{SubSec:e-Model}
\subsubsection{Nonlinearity} \label{SubSubSec:e-Nonl}
The two luminescent processes in LS are scintillation and Cherenkov processes. It is necessary to discuss the contributions to energy nonlinearity from the two effects separately.

\begin{figure}[!htb]
\centering
\includegraphics[width=0.95\hsize]{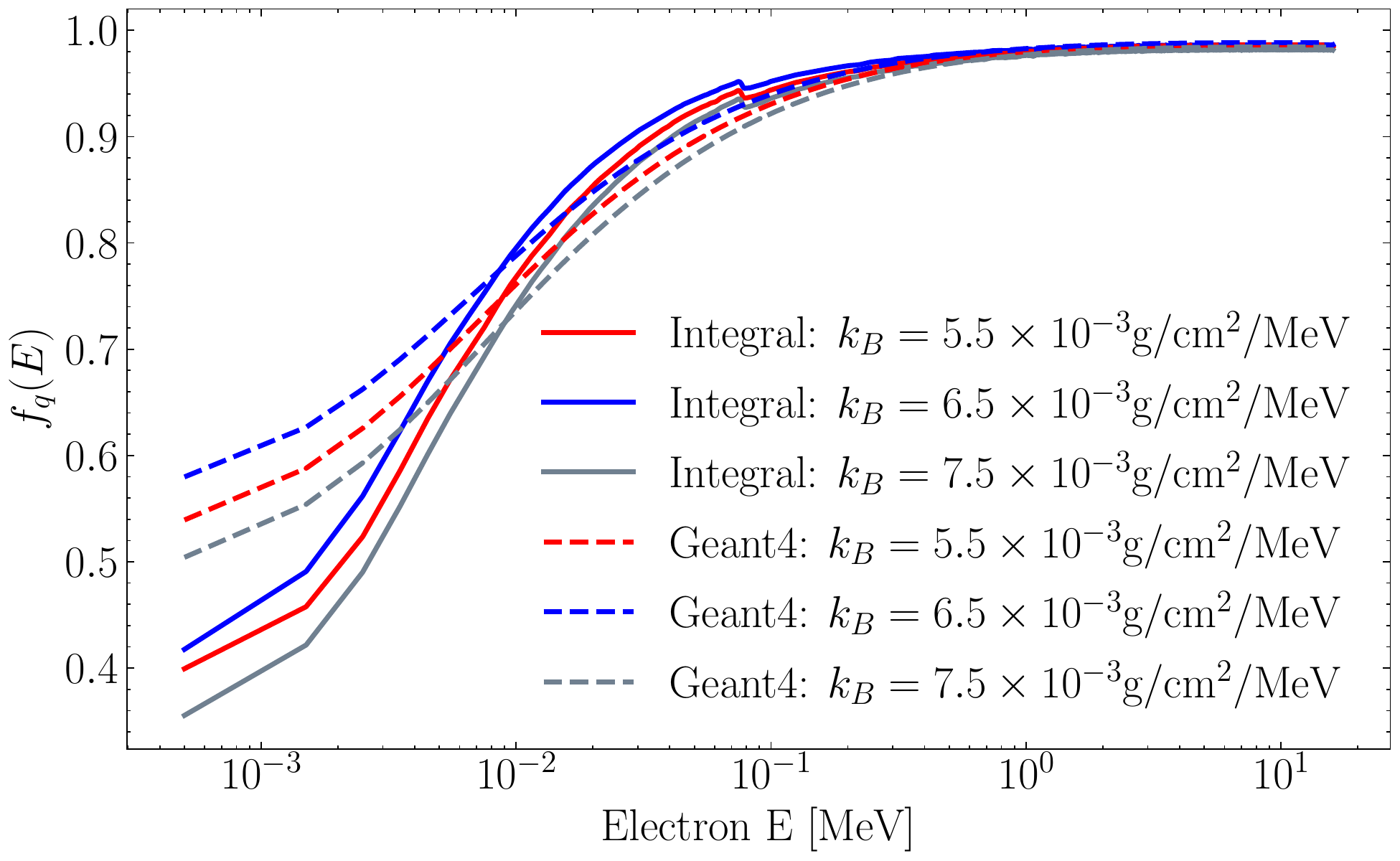}
\caption{The quenching induced nonlinearity as a function of the deposited energy of electrons. Solid lines are derived from the numerical integral with the ESTAR $\mathrm{d}E/\mathrm{d}x$ and dashed lines are Geant4-based simulation results with the production cut of $\SI{0.1}{mm}$. The red, blue and gray lines are nonlinearity curves with $k_\mathrm{B}=5.50\times10^{-3}$g/cm$^{-2}$/MeV, $k_\mathrm{B}=6.50\times10^{-3}$g/cm$^{-2}$/MeV (inherent value in simulation) and $k_\mathrm{B}=7.50\times10^{-3}$g/cm$^{-2}$/MeV, respectively.}
\label{Fig:QuenchNonlinearity} 
\end{figure}

1. \textit{Quenching effect.} Some fraction of the deposited energy is transferred into heat and only the remained quenched energy $E_\mathrm{q}$ is visible by generating photons~\cite{Birks:1964zz}. The quenching induced nonlinearity $f_\mathrm{q}$ is defined as $f_\mathrm{q} = \overline{E_\mathrm{q}}/E$, and the expected value of visible energy from scintillation photons ($\overline{E_\mathrm{s}}$) can be expressed by introducing the scintillation light yield $Y_\mathrm{s}$, which is the number of scintillation PEs per unit visible energy:
\begin{equation}
    \overline{E_\mathrm{s}} = f_\mathrm{q}(k_\mathrm{B}, E) \cdot E \cdot Y_\mathrm{s} / Y,
    \label{Eq:Es}
\end{equation}
where $k_\mathrm{B}$ is the Birks' coefficient and larger $k_\mathrm{B}$ yields larger quenching. Implementations of the quenching effect in Geant4 are based on the Birks' model in discrete steps. Besides, we used a numerical integral of Birks' model with stopping power data from ESTAR~\cite{Berger:1993} for cross-check. The different calculation results of $f_\mathrm{q}(E)$ are shown in figure~\ref{Fig:QuenchNonlinearity}. The curves have apparent discrepancies in the low energy region, while similar best-fit curves of the total nonlinearity and resolution have been obtained in the follow-up analysis, which validates the feasibility and robustness of our model. Hereafter, the results are based on the numerical integral curves. 

\begin{figure}[!htb]
\centering
\includegraphics[width=0.95\hsize]{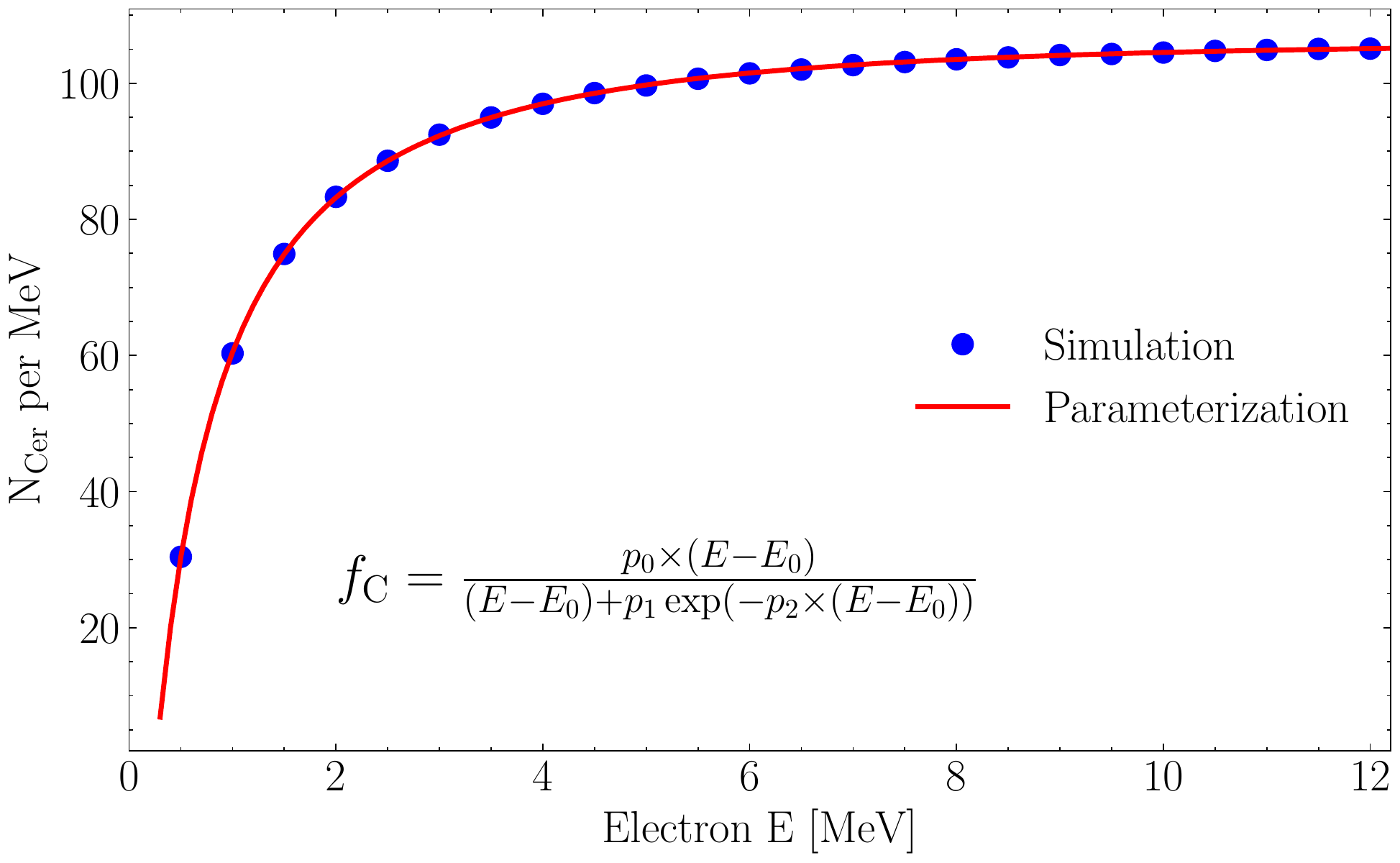}
\caption{The Cherenkov PEs yield $f_\mathrm{C}$ for $e^-$ from simulation (blue markers). The parametrization form is shown as the red line.}
\label{Fig:CerenkovNonlinearity}
\end{figure}

2. \textit{Cherenkov radiation.} Cherenkov effect heavily relies on LS refractive index and the absorption re-emission probability, especially in the ultraviolet region. The Frank-Tamm formula~\cite{Frank:1937fk} is implemented in Geant4 for Cherenkov intensity calculation. The shape of the Cherenkov photoelectron yield as a function of the electron deposited energy $f_\mathrm{C}\left( E \right)$ can be obtained from Geant4 simulation and is shown in figure~\ref{Fig:CerenkovNonlinearity}. To avoid the heavy dependency on the LS optical properties in simulation, an empirical parametrization description of Cherenkov light yield is constructed as,
\begin{equation}
\begin{array}{l}
    f_\mathrm{C}\left( E \right) = \left\{
    \begin{aligned}
    & 0, E^\prime < 0,  \\
    & \frac{p_0\cdot E^\prime}{ E^\prime + p_1\cdot e^{-p_2\cdot E^{\prime}}},  E^{\prime}\geq 0,
    \end{aligned}
    \right.
    \end{array}
    \label{Eq:fCer}
\end{equation}
where $E^\prime = E-E_0$, and $E_0$ is the energy threshold of the Cherenkov effect. The fitting curve is also shown in figure~\ref{Fig:CerenkovNonlinearity}, and there is a good agreement between the simulation and our model.

In summary, the expected visible energy is the sum of the scintillation part and the Cherenkov part, as shown in Eq.~\eqref{Eq:elecNPE},
\begin{align}
    \overline{E_\mathrm{vis}} &= \overline{E_\mathrm{s}} + \overline{E_\mathrm{C}} \nonumber \\
    &= f_\mathrm{q}\left( k_\mathrm{B}, E\right)\cdot E \cdot Y_\mathrm{s} / Y +  f_\mathrm{C}\left(E, \bm{p}\right)\cdot E / Y, 
    \label{Eq:elecNPE}
\end{align}
where the parameter $\bm{p}$ are the four parameters involved in Eq.~\eqref{Eq:fCer}. A six-parameter energy nonlinearity model has been established for $e^-$ based on the physical origins.

\subsubsection{Resolution} \label{SubSubSec:e-Resol}

\begin{figure*}[htbp]
\centering
\subfigure{
\label{Fig:e-ResolDecomp}
\includegraphics[width=0.45\hsize]{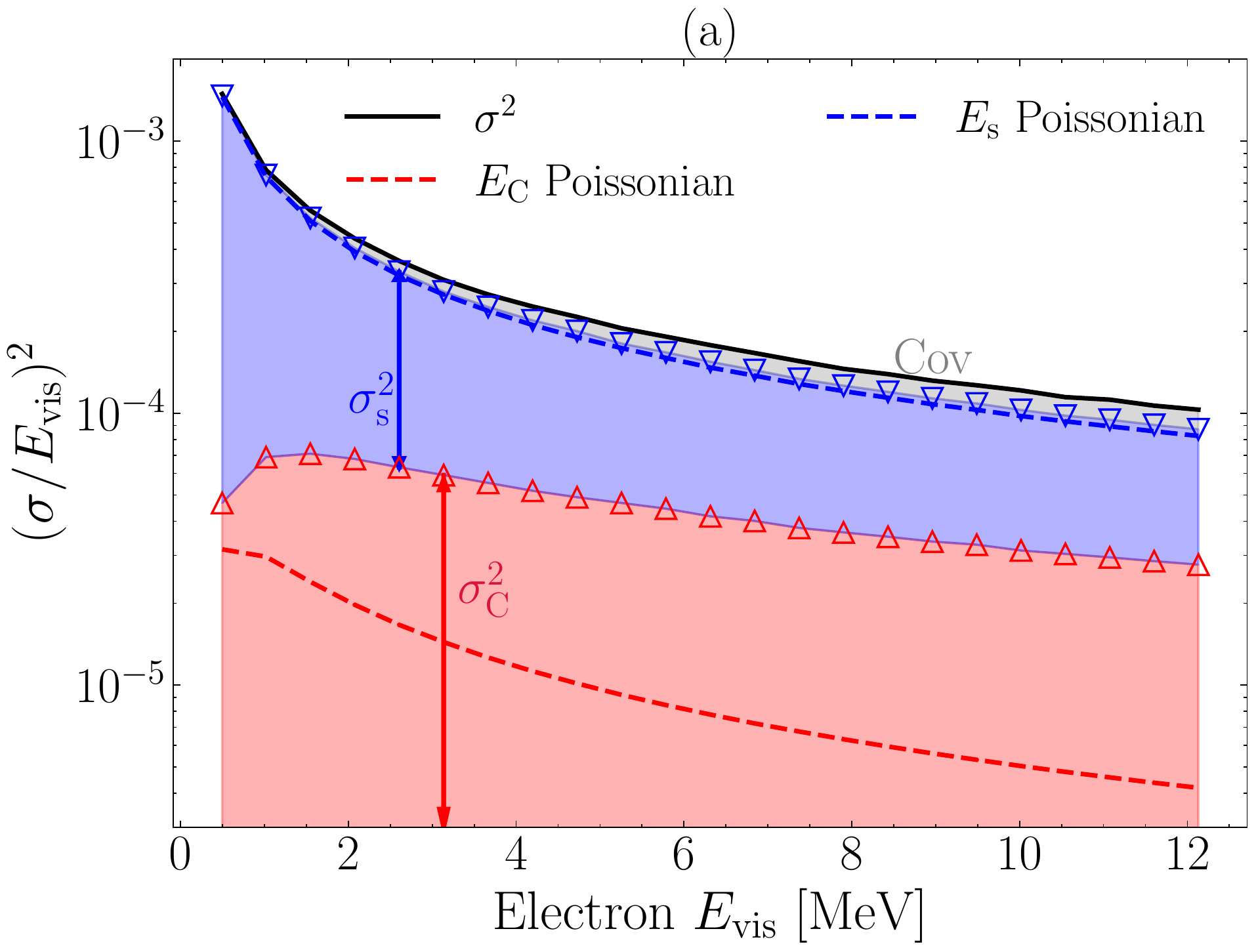}
}
\subfigure{
\label{Fig:e-ResolFit}
\includegraphics[width=0.45\hsize]{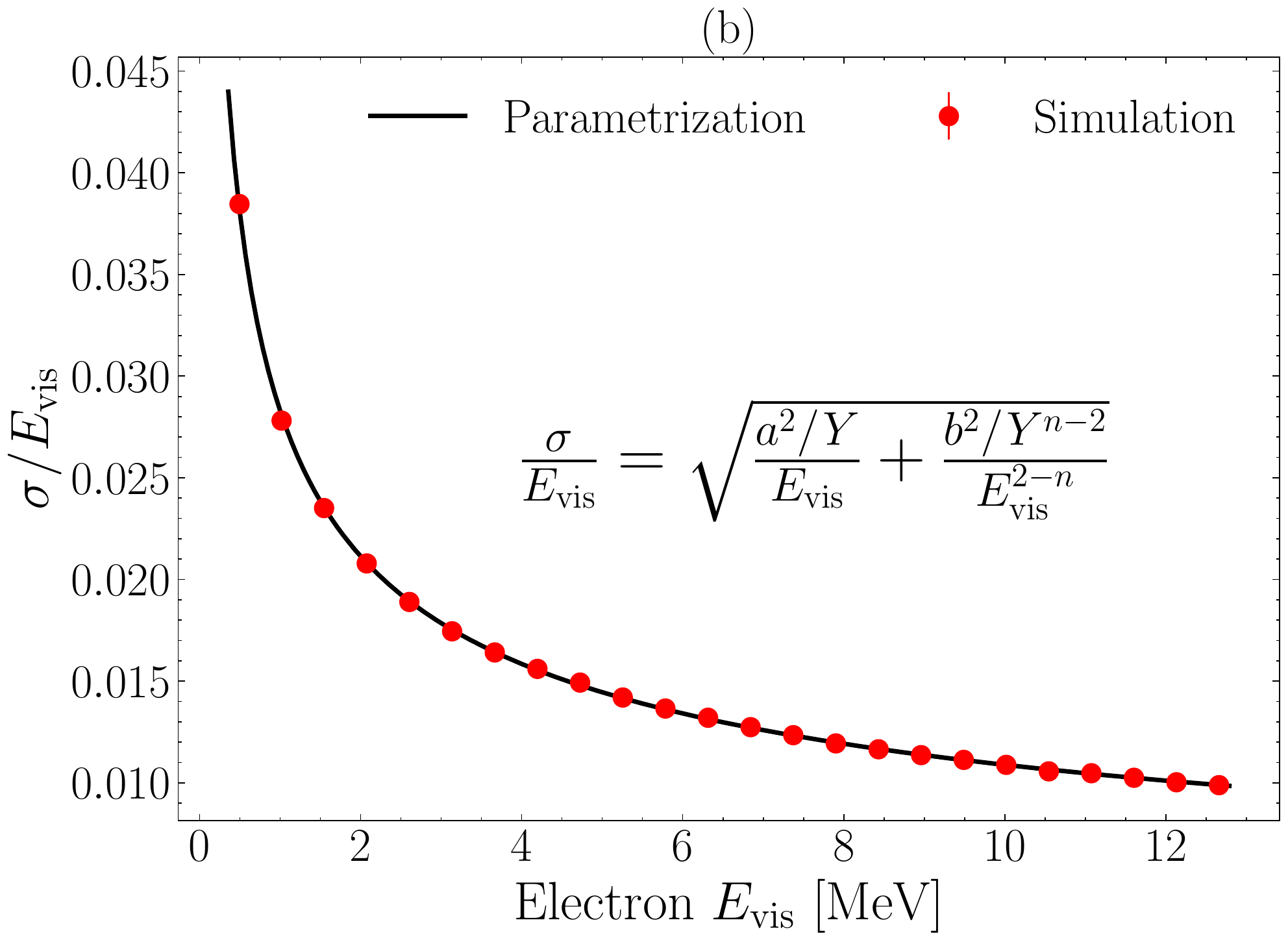}
}
\caption{(a): decomposition of $e^-$ energy resolution. The black line is the total energy resolution from the simulation. The three shadow regions represent the Cherenkov part $(\sigma_\mathrm{C}/E_\mathrm{vis})^2$ (red), scintillation part $(\sigma_\mathrm{s}/E_\mathrm{vis})^2$ (blue) and covariance part $2\cdot \mathrm{Cov}\left[ E_\mathrm{s}, E_\mathrm{C} \right]/E_\mathrm{vis}^2$ (gray). The boundary markers are the Monte Carlo simulation, blue inverted triangles for $E_\mathrm{s}$ and red triangles for $E_\mathrm{C}$. The blue dashed line and red dashed line refer to the Poisson statistics for scintillation and Cherenkov photons. Large excess w.r.t. the Poisson statistics exists in the Cherenkov case. (b): parametrization of the energy resolution of $e^-$ (simulation data as red markers, the best fit of parametrization form as the black line).}
\end{figure*}

Similar to the energy nonlinearity model, the contributions to the energy resolution from scintillation photons and Cherenkov photons have been studied separately. Considering the correlation between $E_\mathrm{s}$ and $E_\mathrm{C}$, the total fluctuation of the visible energy can be decomposed as:
\begin{eqnarray}
\sigma^2 = \sigma_\mathrm{s}^2 + \sigma_\mathrm{C}^2 + 2 \cdot \mathrm{Cov}\left[ E_\mathrm{s}, E_\mathrm{C} \right],
\label{Eq:e-SigmaDecomp}
\end{eqnarray}
where $\sigma_s$ and $\sigma_C$ represent the standard deviation of $E_\mathrm{s}$ and $E_\mathrm{C}$, and $\mathrm{Cov}\left[ E_\mathrm{s}, E_\mathrm{C} \right]$ is the covariance defined as $\mathrm{Cov}\left[ E_\mathrm{s}, E_\mathrm{C} \right] =\overline{E_\mathrm{s}\cdot E_\mathrm{C}}-\overline{E_\mathrm{s}}\cdot\overline{E_\mathrm{C}}$. In figure~\ref{Fig:e-ResolDecomp} the three parts in Eq.~\eqref{Eq:e-SigmaDecomp} have been shown as the three shadow regions. The scintillation process is close to Poissonian as expected (blue inverted triangles). The Cherenkov photons have a rather large smearing due to particle track length fluctuation (red circles). A longer track length leads to more Cherenkov photons and relatively smaller $\mathrm{d}E/\mathrm{d}x$, which induces a smaller quenching effect. Thus, a positive correlation between $E_\mathrm{s}$ and $E_\mathrm{C}$ degrades the energy resolution furthermore (gray region). The energy resolution for LS detectors is sensitive to the photon composition, and a higher Cherenkov ratio yields worse resolution with the same photon statistics. 

Due to the strong correlation, decoupling of the three terms in Eq.~\eqref{Eq:e-SigmaDecomp} is difficult without separate bench measurements. Instead, the energy resolution for LS detectors is often empirically modeled as 
\begin{eqnarray}
R^2 & = & \left(\frac{\sigma}{\overline{E_\mathrm{vis}}}\right)^2 =\frac{a^2 / Y}{\overline{E_\mathrm{vis}}} + \frac{b^2 \left( Y\right)^{n-2}}{\left( \overline{E_\mathrm{vis}}\right)^{2-n}} + \frac{c^2 / Y^2 }{\overline{E_\mathrm{vis}}^2}, \label{Eq:e-ResolModel}\\
    & = & R_{\mathrm{stat}}^2 + R_{\mathrm{non-Pois}}^2 + R_{\mathrm{noise}}^2, \label{Eq:e-ResolDecomp}
    \label{Eq:elecResolE}
\end{eqnarray}
where the three terms in Eq.~\eqref{Eq:e-ResolModel} are redefined as the corresponding terms in Eq.~\eqref{Eq:e-ResolDecomp}. The $a$-related term is mainly induced by statistical fluctuation, the $b$-related term is dominated by a non-Poisson fluctuation of Cherenkov PEs and $c$ is caused by PMT dark noises. By the insertion of $Y$ in Eq.~\eqref{Eq:e-ResolModel}, $a, b$ and $c$ are all dimensionless variables. The parameter $n$ in Eq.~\eqref{Eq:e-ResolModel} is often fixed as $2$ in previous studies~\cite{JUNO:2020xtj,DayaBay:2019fje,DayaBay:2016ggj}, while it is released as a free parameter in this analysis for more flexible description extending to the higher energy region.

Suppose that PMT dark noises in the readout window obey Poisson statistics, the noise term is particle independent and can be estimated as
\begin{equation}
    c^2 = R_\mathrm{DN} \times T \times N_\mathrm{PMT},
    \label{Eq:cterm}
\end{equation}
with the total number of PMTs $N_\mathrm{PMT}$, average dark noise rate $R_\mathrm{DN}$ and readout time window length $T$. There will be a trade-off between less dark noises in shorter windows and more signal photons in longer windows during the optimization of energy reconstruction, which requires a separate study. In the followings, the $R_\mathrm{noise}$ term is neglected as no PMT noises are considered. But it is straightforward to include dark noises in the future by introducing $c$-related term into the model. figure~\ref{Fig:e-ResolFit} shows the resolution of $e^-$ from simulation, also the parametrization form based on Eq.~\eqref{Eq:e-ResolModel} is plotted together.

\subsection{Energy response model of \texorpdfstring{$\boldsymbol{\gamma}$}{gamma}}\label{SubSec:gammaModel}
The connections between energy resolution of $\gamma$ and $e^-$ have been constructed in section~\ref{Sec:connections}. In Eq.~\eqref{Eq:gamNonl} and Eq.~\eqref{Eq:gamRes}, the calculation of energy response for mono-energetic $\gamma$s requires applying $e^-$ nonlinearity (Eq.~\eqref{Eq:elecNPE}) and resolution (Eq.~\eqref{Eq:e-ResolModel}) on the primary $e^\pm$ collections of all deposition modes. To avoid complex modeling, the method adopted here is to sample deposition modes from simulation based on a certain physical model. With sufficient sampling statistics, the calculation results are supposed to converge to the true values.

\begin{figure*}
	\centering
	\includegraphics[width=0.9\textwidth]{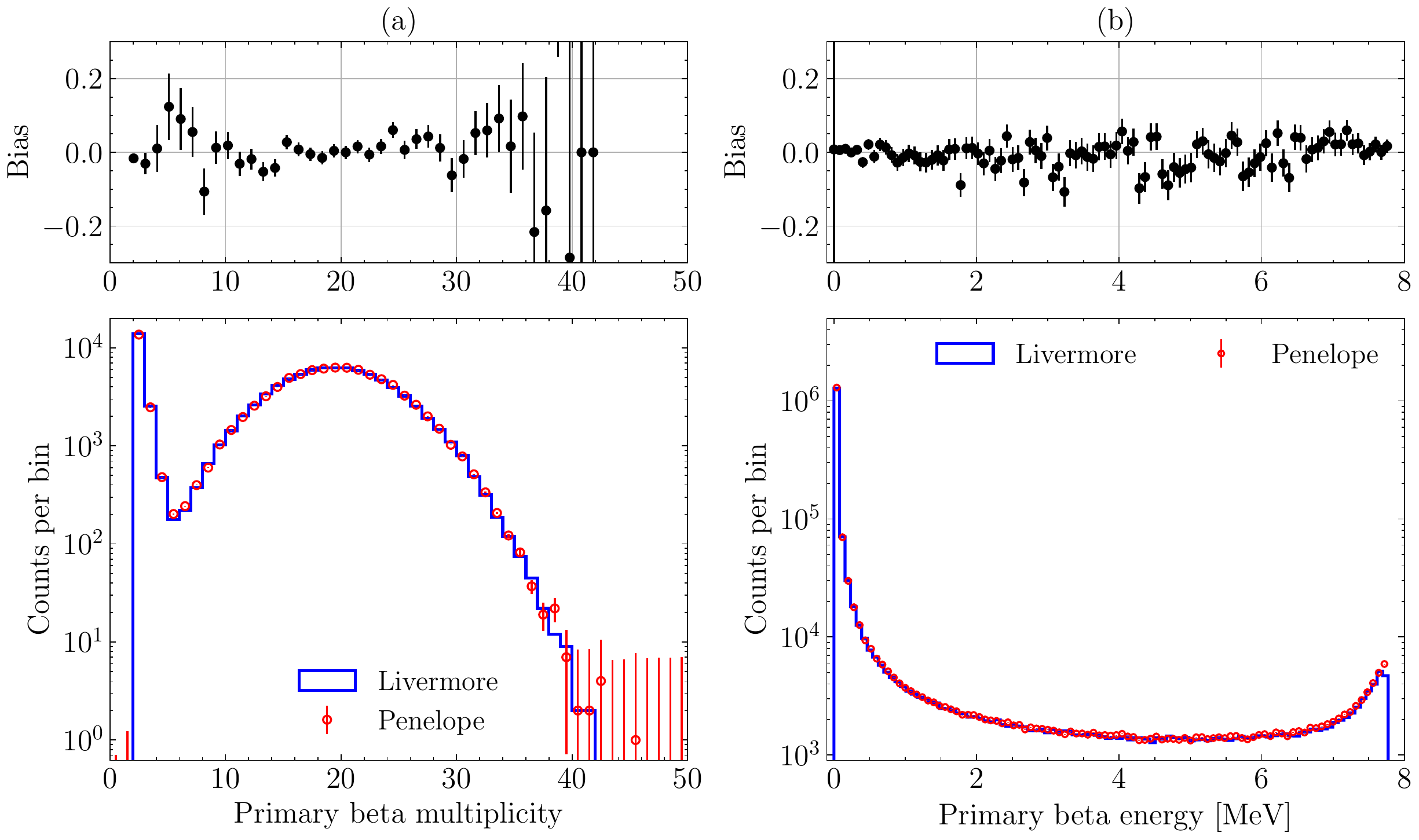}
	\caption{Multiplicity distribution (a) and energy distribution (b) for $\SI{8}{MeV}~\gamma$s with Livermore model (blue) and Penelope model (red). The relative bias is defined as $(L-P)/{L}$ where $L$ for the Livermore model and $P$ for the Penelope model, and displayed in the upper panels.}
	\label{Fig:modelVal}
\end{figure*}

To test the dependency on physical models of energy deposition of $\gamma$s, two common low-energy electromagnetic models in Geant4, namely Livermore and Penelope, have been compared. It is claimed by the Geant4 group that these two models can provide reliable results covering electrons and photons physics from $250\,\mathrm{eV}$ to $1\,\mathrm{GeV}$. Taking $8\,\mathrm{MeV}$ $\gamma$s as an example, consistent results of the energy and multiplicity distributions of primary $e^\pm$ have been obtained between the two models in figure~\ref{Fig:modelVal}. Besides, the interaction cross sections for $\gamma$ energy deposition are well theoretically calculated and measured in ref.~\cite{Berger2019}. An algorithmic calculation of primary $e^\pm$ distributions has been developed as in ref.~\cite{Kampmann:2020hto} and has excellent agreements with Geant4 simulation. Based on the above cross-validations on the $\gamma$s interactions in the reactor antineutrino energy range, we choose the Livermore model in the following simulation.

\begin{figure}[!htb]
	\centering
	\includegraphics[width=0.95\hsize]{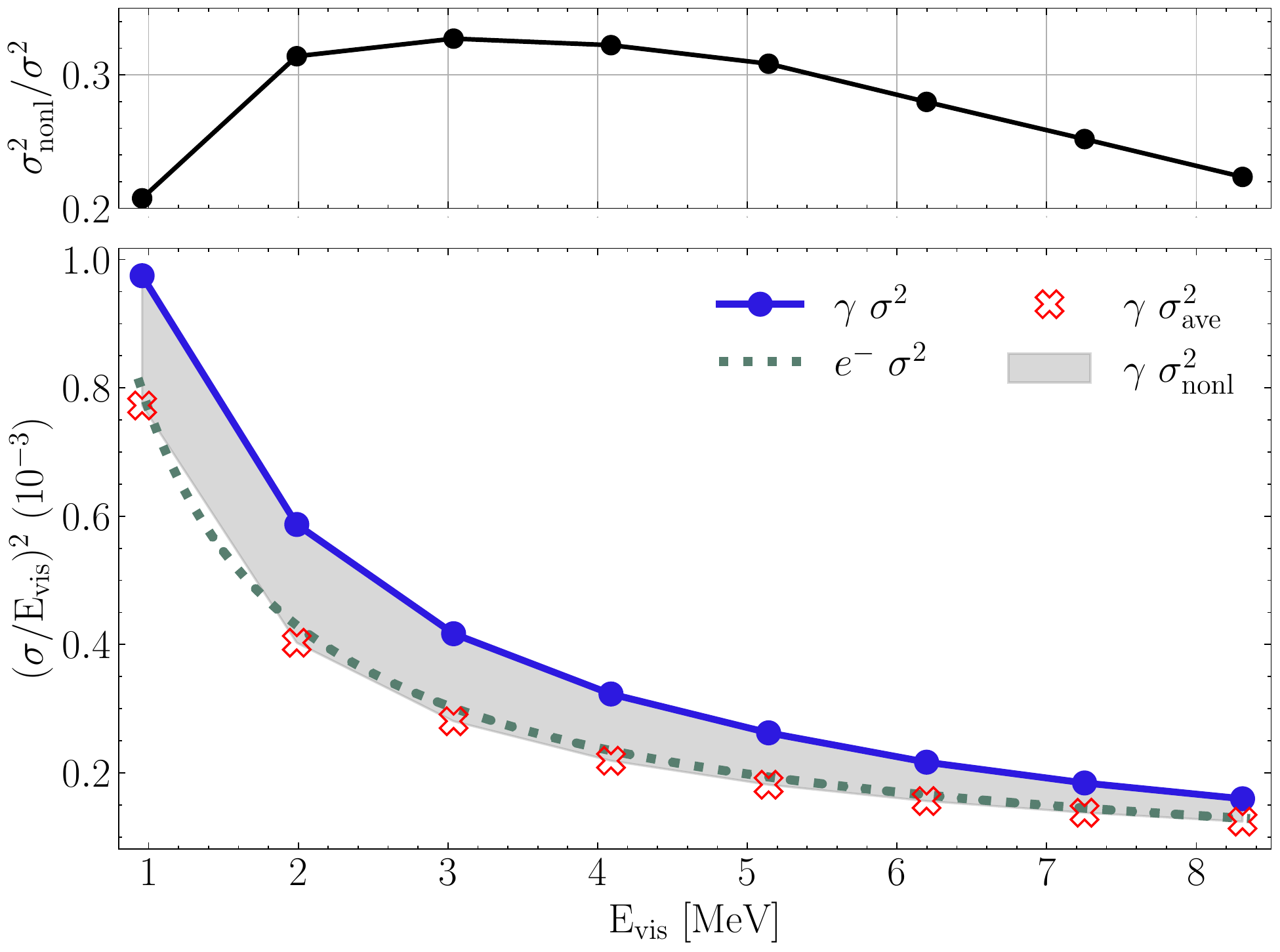}
	\caption{Lower panel: decomposition of $\gamma$ energy resolution according to Eq.~\eqref{Eq:gamResDecomp}, with $\sigma^2$ in blue solid line with round markers and $\sigma_\mathrm{ave}^2$ in red cross markers. The $e^-$ energy resolution curve is shown as the dark green dotted line for comparison. The worse resolution for $\gamma$ than $e^-$ comes from $\sigma_\mathrm{nonl}^2 = \sigma^2 - \sigma_\mathrm{ave}^2$ (gray region) which is caused by the dispersion among deposition modes. The remained $\sigma_\mathrm{ave}^2$ is slightly smaller than $\left[ \sigma^{e^-}\right]^2$ due to less Cherenkov photons. Upper panel: the ratio of $\sigma_\mathrm{nonl}^2/\sigma^2$, which is larger than $20\%$ below $10\,\mathrm{MeV}$.}
	\label{Fig:gammaDecomp}
\end{figure}

Decomposition of $\gamma$ energy resolution as Eq.~\eqref{Eq:gamResDecomp} is shown in figure~\ref{Fig:gammaDecomp}. The energy resolution of $e^-$ is also displayed as the dark green dotted line for comparison. The excess of $\gamma$ resolution compared with $e^-$ mainly comes from the nonlinearity induced smearing term $\sigma_\mathrm{nonl}^2$, which contributes more than $20\%$ in the energy range below $10\,\mathrm{MeV}$. The remained term $\sigma_\mathrm{ave}^2$ (red cross markers) is slightly smaller than $e^-$ resolution with the same visible energy, because lower energy $e^\pm$ in $\gamma$ energy deposition has fewer Cherenkov photons.

\subsection{Energy response model of \texorpdfstring{$\boldsymbol{e^+}$}{e+}}\label{SubSec:e+Model}
Precision measurement of reactor neutrino oscillations requires a deep understanding of the energy resolution of $e^+$.
Meanwhile, knowledge of the resolution contributors will provide guidelines for future improvement. As discussed in section~\ref{Sec:connections}, the energy response of $e^+$ in LS can be conveniently expressed with the energy response of $e^-$ and calibration source $^{68}$Ge (Eq.~\eqref{Eq:e+NonlModel} and Eq.~\eqref{Eq:e+ResModel}). Therefore, by adding calibration data of $^{68}$Ge upon Eq.~\eqref{Eq:elecNPE} and Eq.~\eqref{Eq:e-ResolModel}, the detailed expressions of $e^+$ energy resolution can be written as, 
\begin{align}
    f^{e^+} &= \frac{\overline{E_\mathrm{vis}^{e^-}}(T) + \overline{E_\mathrm{vis}^{^{68}\mathrm{Ge}}} }{T+2m_e} \nonumber \\
    &= \frac{\left[ f_\mathrm{q}(k_\mathrm{B}, T)\cdot T \cdot Y_\mathrm{s}/Y + f_\mathrm{C}(T, \mathbf{p})\cdot T/Y \right] + \overline{E_\mathrm{vis}^{^{68}\mathrm{Ge}}}}{T + 2m_e}, \label{Eq:e+NonlDetailed} \\
    R^{e^+} &= \frac{\sqrt{ \left[a^2/Y\cdot \overline{E_\mathrm{vis}^{e^-}}(T) + b^2Y^{n-2}\cdot (\overline{E_\mathrm{vis}^{e^-}}(T))^n\right] + \left[\sigma^{{^{68}\mathrm{Ge}}}\right]^2}}{\overline{E_\mathrm{vis}^{e^-}}(T) + \overline{E_\mathrm{vis}^{^{68}\mathrm{Ge}}}}, \label{Eq:e+ResDetailed}
\end{align}

where $\overline{E_\mathrm{vis}^{^{68}\mathrm{Ge}}}$ and $\sigma^{{^{68}\mathrm{Ge}}}$ represent the expected value and standard deviation of the calibrated visible energy spectrum of $^{68}$Ge.

\section{Model tuning and results} \label{Sec:FitResults}
\subsection{Model inputs} \label{SubSec:MCInputs}
The radioactive gamma sources are the most important for LS detector calibration. In our Monte Carlo simulation, the gamma sources from ref.~\cite{JUNO:2020xtj} are used, as shown in table~\ref{Tab:CalibGamma}. For simplicity, all $\gamma$ sources are considered as bare sources without enclosures and simulated at the detector center. To reach $0.01\%$ statistical uncertainty of the energy peaks, $\SI{50}{k}$ events for each source are generated. 
\begin{table*}[t]
\centering
\caption{List of the calibration sources from~\cite{JUNO:2020xtj}.}
\label{Tab:CalibGamma}       
\begin{tabular}{lcr}
\hline
Sources/Processes & Type & Radiation (unit: MeV) \\
\hline
$^{137}$Cs              & $\gamma$      & $0.662$ \\
$^{54}$Mn               & $\gamma$      & $0.835$ \\
$^{60}$Co               & $\gamma$      & $1.173+1.333$ \\
$^{40}$K                & $\gamma$      & $1.461$  \\
$^{68}$Ge               & $e^+$         & annihilation $0.511+0.511$ \\
$^{241}$Am-Be           & $n$, $\gamma$ & $n$ + $4.43$ ($^{12}$C$^*$)  \\
(n, $\gamma$)$^{12}$C   & $\gamma$      & $4.94 \left( 68\%\right)$ or $3.68+1.26 (32\%)$ \\
$^{241}$Am-$^{13}$C     & $n$, $\gamma$ & $n$ + $6.13$ ($^{16}$O$^*$) \\
(n, $\gamma$)H          & $\gamma$      & $2.22$ \\
\hline
\end{tabular}
\end{table*}

Besides the regular gamma calibration sources, some isotopes with continuous $\beta$ spectra induced by energetic cosmic muons can also serve as the model inputs. One important source is $^\mathrm{12}\mathrm{B}$, which decays with a $Q$ value about $\SI{13.4}{MeV}$ and lifetime around $\SI{29}{ms}$. The visible energy spectrum of $^\mathrm{12}\mathrm{B}$ spans from $\SI{0}{MeV}$ to around $\SI{14}{MeV}$. 
Another useful sample is Michel $e^-$ generated by stopped cosmic muons. Michel $e^-$ has a wide energy spectrum with the cut-off energy at $\SI{52.8}{MeV}$. To achieve relatively low statistical uncertainties, the $^\mathrm{12}\mathrm{B}$ samples and Michel $e^-$ events are assumed as $\SI{100}{k}$ and $\SI{400}{k}$ respectively in this study. 
Taking the JUNO detector as an instance, where the muon rate is estimated as $\SI{3.6}{Hz}$~\cite{JUNO:2021kxb}, the charge ratio is approximated as the value at the sea level~\cite{Guo:2020orw} and the muon stopping rate is around $4\%$, the statistics is equivalent to three month running period approximately. All the isotope spectra are simulated at the detector center without considering the detector geometric effects.

\subsection{Statistical approach via \texorpdfstring{$\boldsymbol{\chi^2}$}{chi2} minimization}
\label{SubSec:StatMethods}
The inputs for model fitting are the energy spectra of sources mentioned in section~\ref{SubSec:MCInputs}. The whole energy spectra of $\gamma$ sources are utilized. For $^\mathrm{12}\mathrm{B}$ spectrum, an energy window from $3\,\mathrm{MeV}$ to $12\,\mathrm{MeV}$ is chosen, both for natural radioactivity suppression and covering the high energy range of reactor antineutrinos.

To determine the parameters in our energy model, a $\chi^2$ statistics is defined as Eq.~\eqref{Eq:chi2} by comparing predictions (P) and measurements (M) of energy spectra for all input sources, 
\begin{eqnarray}
\chi^2(Y_\mathrm{s}, k_\mathrm{B}, \bm{p}, a, b, n) 
& = & \sum_i \left[ \sum_{j^i} \frac{ \left(P_{j^i}^\gamma - M_{j^i}^\gamma \right)^2}{\left( \sigma_{j^i}^\gamma \right)^2} \right] \nonumber \\
& + & \sum_k \left[ \frac{\left(P_k^{^\mathrm{12}\mathrm{B}} - M_k^{^\mathrm{12}\mathrm{B}}\right)^2}{\left(\sigma_k^{^\mathrm{12}\mathrm{B}}\right)^2} \right].
\label{Eq:chi2}
\end{eqnarray}
The first term in Eq.~\eqref{Eq:chi2} is for $\gamma$s. The summation of $i$ enumerates all calibration $\gamma$ sources, and the summation of $j^i$ enumerates all energy bins for the $i$-th source. $M_{j^i}$ and $\sigma_{j^i}$ are the count of $j$-th energy bin and its statistical uncertainty. Neglecting the energy leakage in large LS detectors via a fiducial volume, the ideal energy spectra of $\gamma$ sources are approximately expected as normal distributions $\mathcal{N}\left( \overline{E_\mathrm{vis}^\gamma}, \sigma^\gamma\right)$. Thus, the count of $j$-th energy bin ($P_{j^i}$) can be predicted according to $\overline{E_\mathrm{vis}^\gamma}$ and $\sigma^\gamma$, where the $\overline{E_\mathrm{vis}^\gamma}$ and $\sigma^\gamma$ can be calculated using the $e^-$ energy response. The second term in Eq.~\eqref{Eq:chi2} is the spectrum comparison of $^\mathrm{12}\mathrm{B}$. The summation of $k$ enumerates all fitting energy bins of the $^\mathrm{12}\mathrm{B}$ energy spectrum. $M_k^{^\mathrm{12}\mathrm{B}}$ and $\sigma_k^{^\mathrm{12}\mathrm{B}}$ are the count of $k$-th energy bin and its statistical uncertainty. The $^\mathrm{12}\mathrm{B}$ energy spectrum ($P_k^{^\mathrm{12}\mathrm{B}}$) is predicted by first applying the $e^-$ nonlinearity (Eq.~\eqref{Eq:elecNPE}) on the theoretical energy spectrum, and then smearing the spectrum according to the $e^-$ energy resolution function (Eq.~\eqref{Eq:e-ResolModel}). The Michel $e^-$ spectrum is not used as a fitting input considering uncertainties, more details are discussed in section~\ref{SubSec:inputs}.

The description of $e^-$ nonlinearity introduces scintillation PEs yield $Y_\mathrm{s}$, Birks' coefficient $k_\mathrm{B}$ in the quenching effect and $4$ parameters in the Cherenkov effect $f_\mathrm{C}$. Besides, the $e^-$ resolution model requires $3$ more parameters $a, b$ and $n$. Therefore, total $9$ physical parameters are involved in the model fitting. The $\chi^2$ in Eq.~\eqref{Eq:chi2} is minimized with TMinuit.

\subsection{Fitting results}\label{SubSec:FitResults}
\begin{figure*}[!htb]
	\centering
	\subfigure{
    \label{Fig:GammaFitNonl}
    \includegraphics[width=0.42\hsize]{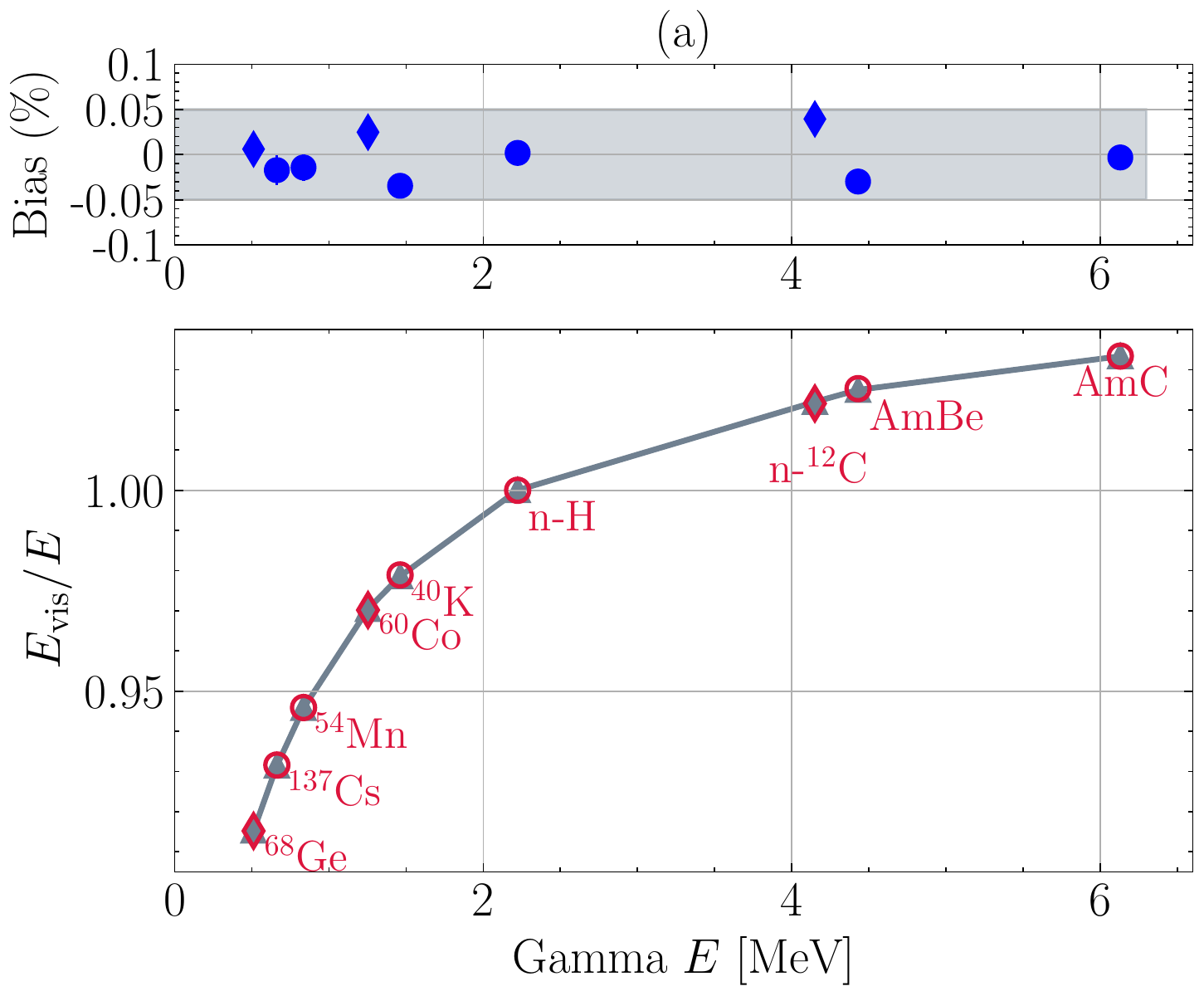}
    }
    \subfigure{
    \label{Fig:GammaFitRes}
    \includegraphics[width=0.42\hsize]{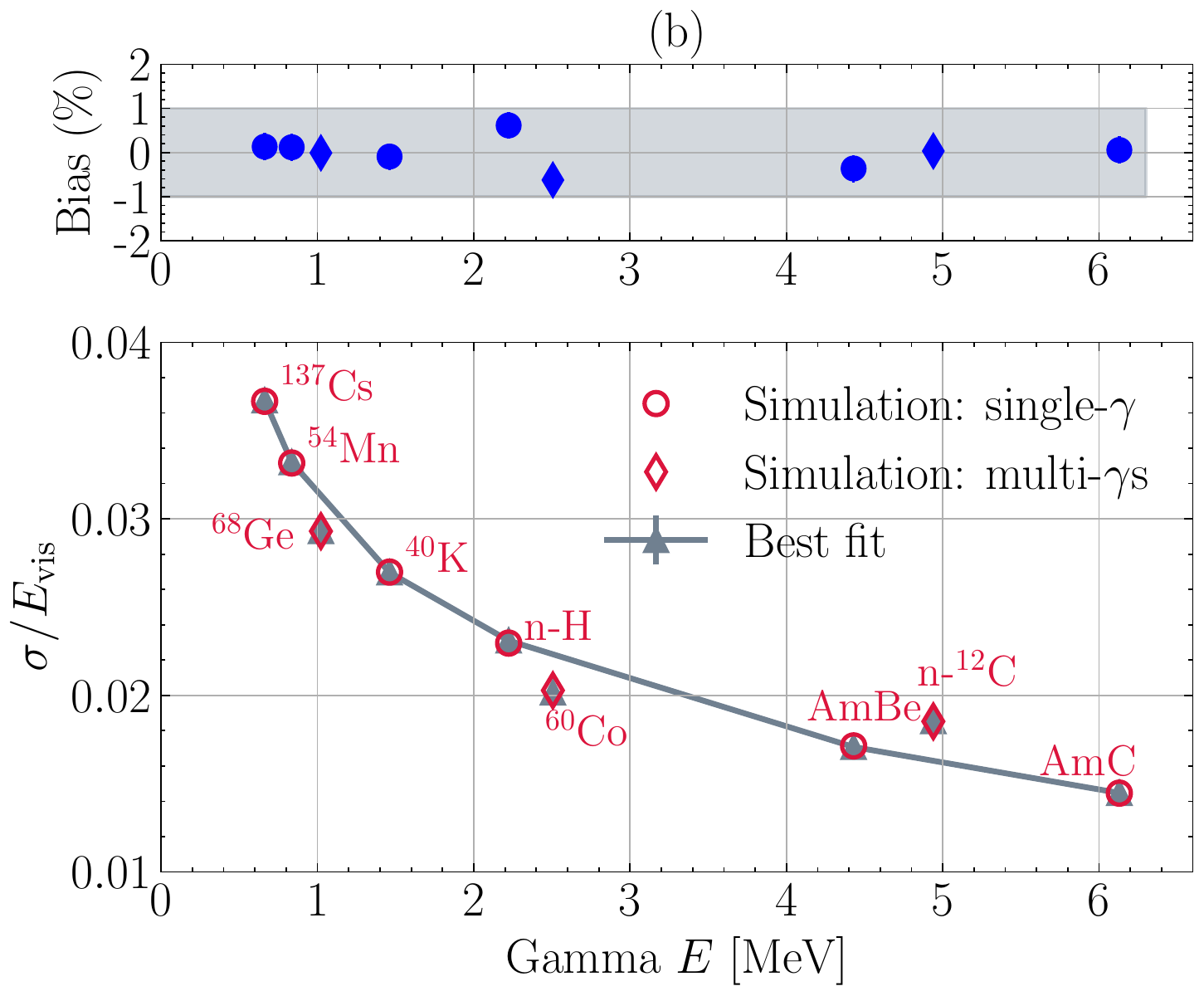}
    }
    \subfigure{
    \label{Fig:B12Fit}
    \includegraphics[width=0.90\hsize]{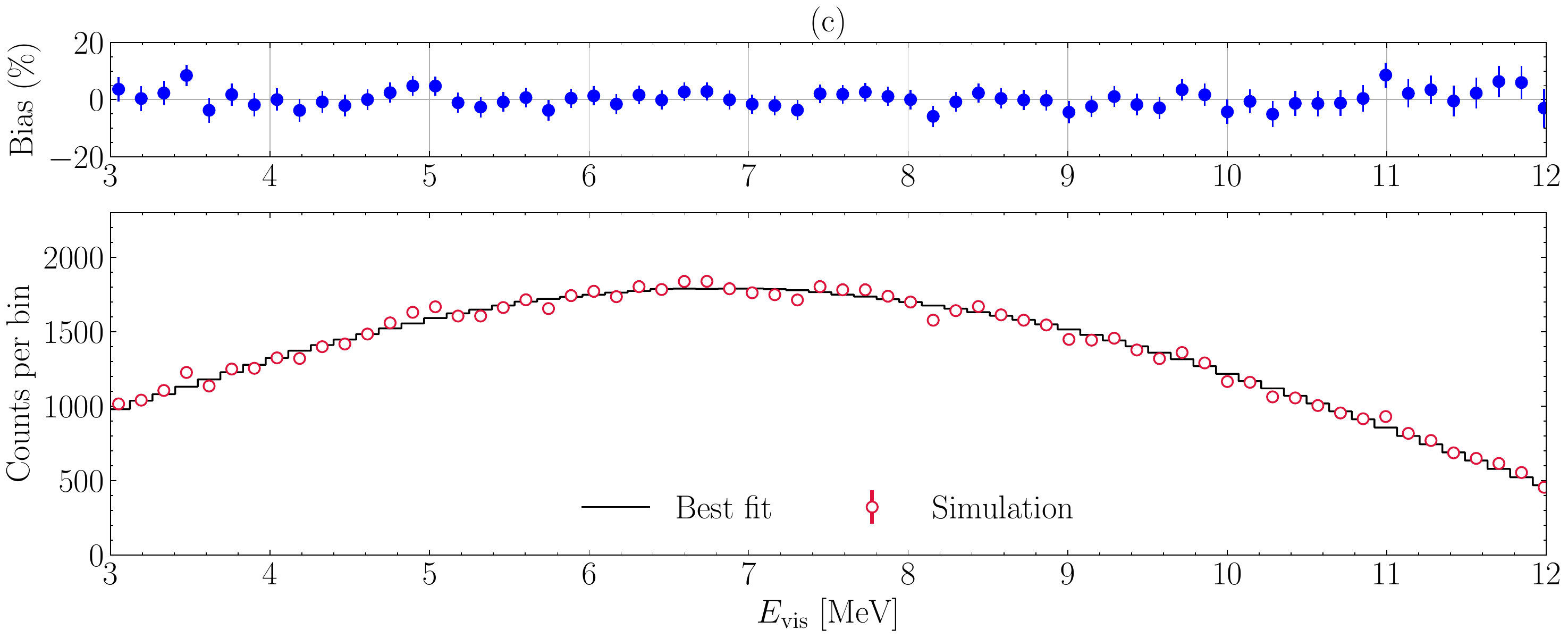}
    }
	\caption{
	Comparison of $\gamma$ energy response and $^\mathrm{12}\mathrm{B}$ energy spectrum between the simulated results (red markers) and the best-fit model (black lines) ((a): gamma energy nonlinearity, (b): energy resolution, (c) $^\mathrm{12}\mathrm{B}$ visible energy spectrum). The relative residual bias, defined as $(F-T)/T$ ($T$ is the simulation value, $F$ is the best-fit model prediction), is within $0.05\%$ and $1\%$ for nonlinearity and resolution, respectively. For nonlinearity of each multi-$\gamma$ source, like $^{60}$Co, $^{68}$Ge and n$^{12}$C, the average energy of all its $\gamma$s is displayed. The resolution of multi-$\gamma$ sources deviates from that of the single-$\gamma$ source, mainly due to different Cherenkov PEs ratios. 
	}
	\label{Fig:fitRes}
\end{figure*}

The fitting procedures have been proceeded in the four simulation configurations mentioned in section~\ref{Sec:MCSim}, and the detector with the energy resolution around $2.9\%$ are demonstrated in this section, where the value of $Y$ is determined as $1400/\mathrm{MeV}$. The fitted nonlinearity and resolution values for all these calibration $\gamma$s are compared with those from simulation in figure~\ref{Fig:GammaFitNonl}-\ref{Fig:GammaFitRes}. The residual bias for nonlinearity and resolution of $\gamma$s can achieve within $0.05\%$ and $1\%$, respectively. As a comparison, the nonlinearity-only calibration in ref.~\cite{JUNO:2020xtj} adopts a $4$-parameter empirical formula and gives a consistent residual bias around $0.1\%$ for $\gamma$ sources.  Among all $\gamma$ calibration sources, there are three multi-$\gamma$s sources, $^{60}$Co, $^{68}$Ge and n$^{12}$C, which have slightly different behavior of energy response compared to the other single $\gamma$ sources. 
The displayed energy $E$ for multi-$\gamma$s sources in figure~\ref{Fig:GammaFitNonl} is calculated as the average value of all the $\gamma$s, while the sum of all the $\gamma$s are displayed for multi-$\gamma$s sources in figure~\ref{Fig:GammaFitRes}.

The $^{68}$Ge and $^{60}$Co sources have better energy resolution compared with single $\gamma$ sources with same deposited energy, because the Cherenkov PEs are less for two softer $\gamma$s, resulting in smaller non-Poisson fluctuation. While for n-$^{12}$C, the worse resolution originates from the $E_\mathrm{vis}$ discrepancy for the two branches (see table~\ref{Tab:CalibGamma}). Furthermore, the best-fit $^\mathrm{12}\mathrm{B}$ spectrum has great consistency with the simulation data (see figure~\ref{Fig:B12Fit}). 

\begin{figure}[!htb]
	\centering
    \includegraphics[width=0.95\hsize]{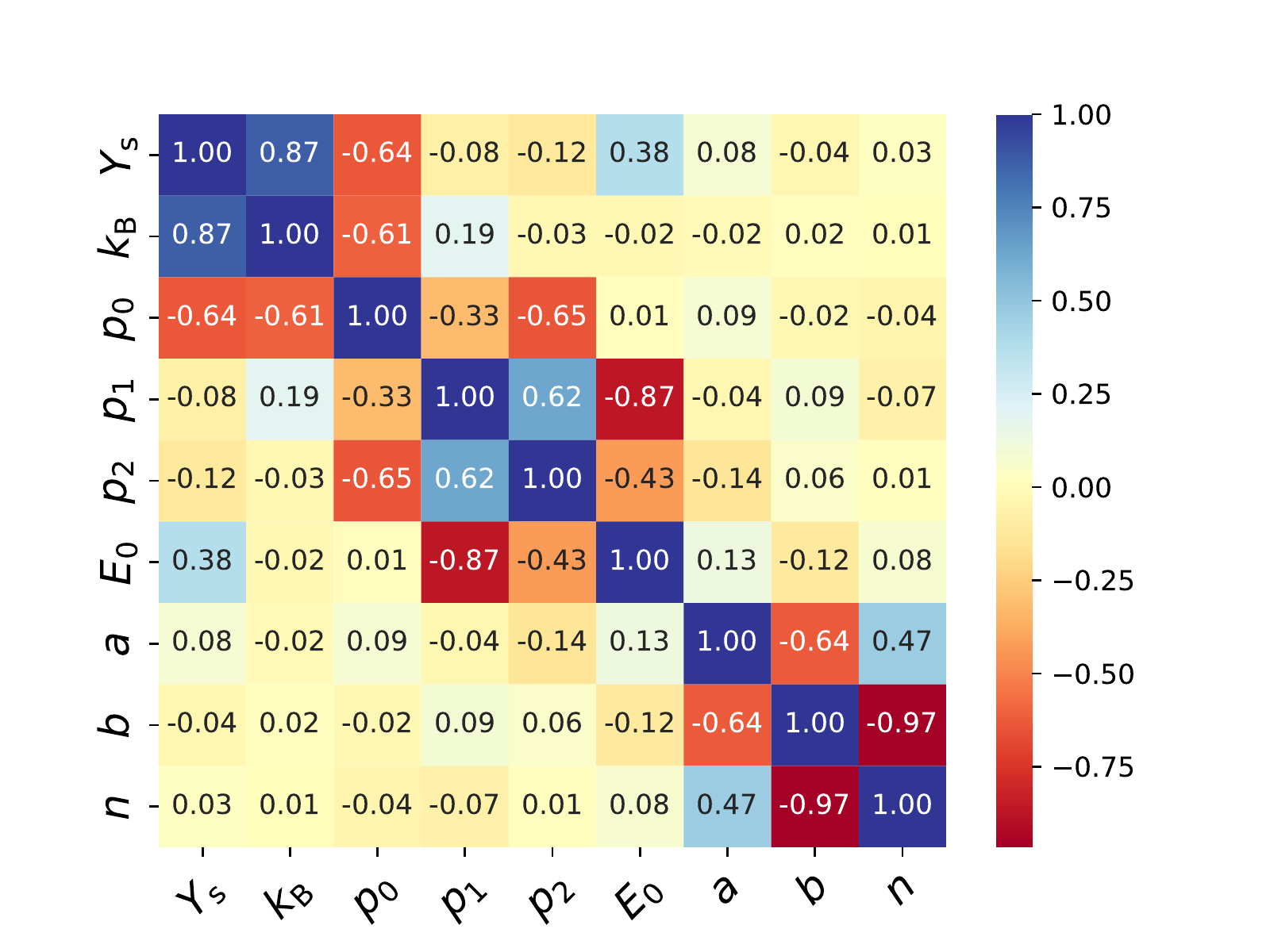}
	\caption{Correlation coefficients of the 9 parameters.}
	\label{Fig:CorrMat}
\end{figure}

The best-fit parameters are listed in table \ref{Tab:FitParam}, and the correlation matrix is shown in figure~\ref{Fig:CorrMat}. The correlation matrix manifests features of the block matrix, where parameters of the scintillation nonlinearity, Cherenkov nonlinearity and energy resolution are strongly correlated internally. The nonlinearity part and the resolution part are only weakly correlated. However, around $60\%$ correlation is found between the amplitudes of scintillation and Cherenkov components ($Y_\mathrm{s}$ and $p_0$), which makes the determination of parameters is entangled and may have bias compared to the nominal settings. For example the best-fit value of the Birks' coefficient $k_\mathrm{B} = (5.76\pm0.03)\times 10^{-3}\,\mathrm{g/cm^2/MeV}$ deviates from the nominal value $6.5\times 10^{-3}\,\mathrm{g/cm^2/MeV}$, while the total nonlinearity fitting performs well as in figure~\ref{Fig:GammaFitNonl}. To disentangle the strong correlation, standalone measurements on different components are important.

\begin{table*}[h!]
    \begin{center}
        \caption{Fitting parameters summary (best-fit values and uncertainties) from $\chi^2$ minimization.}
        \label{Tab:FitParam}
        \begin{tabular}{c|c|c|c|c} 
        \hline
        Parameter & Definition & Unit & Best-fit & Uncertainty \\
        \hline
        $Y_\mathrm{s}$ & scintillation PEs yield & $\rm{MeV^{-1}}$ & $1403.29$            & $0.54$  \\
        $k_\mathrm{B}$ & Birks' coefficient & $ \rm{g/cm^2/MeV} $    & $5.76\times 10^{-3}$ & $ 0.03\times10^{-3} $   \\
        \hline
        $p_0$   & \multirow{4}{7em}{parameters in $f_\mathrm{C}$ (Eq.~\eqref{Eq:fCer})} &  & $91.98$ & $0.60$  \\ 
        $p_1$ & & $\mathrm{MeV}$ & $0.556$ & $0.023$  \\
        $p_2$ & & $\rm{MeV^{-1}}$ & $0.277$ & $0.029$  \\
        $E_0$ & & $\mathrm{MeV}$ & $0.192$ & $0.007$  \\
        \hline\hline
        $a$   & \multirow{3}{6em}{parameters in $R$ (Eq.~\eqref{Eq:e-ResolModel})} &                  & $9.74\times 10^{-1}$ & $3.1\times 10^{-3}$  \\  
        $b$   & &                  & $4.51\times 10^{-2}$ & $2.80\times10^{-3}$  \\
        $n$   & &                  & $1.62$               & $0.01$  \\
        \hline
        \end{tabular}
    \end{center}
\end{table*}

\subsection{\texorpdfstring{$\boldsymbol{e^+}$}{e+} nonlinearity and resolution from data-driven calibration}
\begin{figure*}[!htb]
	\centering
	\includegraphics[width=0.9\textwidth]{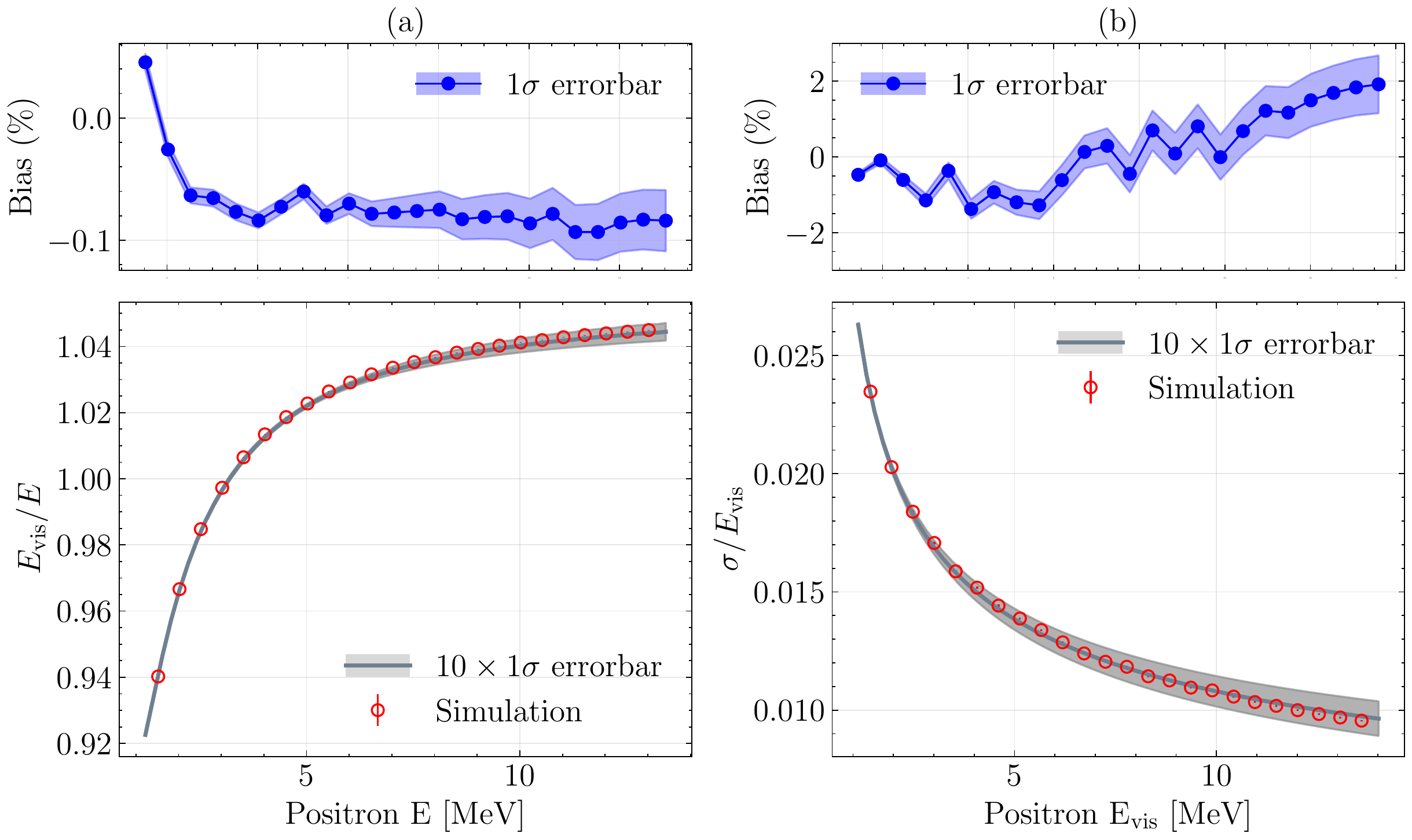}
	\caption{
	Comparison of $e^+$ energy response between the prediction and the simulated data ((a): nonlinearity, (b): resolution). The relative residual bias, with the same definition in figure~\ref{Fig:fitRes}, is less than $0.1\%$ and $2\%$ for nonlinearity and resolution, respectively. The uncertainty bands are scaled for better visualization.
	}
	\label{Fig:e+Results}
\end{figure*}

To validate the model fitting, mono-energetic $e^+$ samples are simulated at the detector center covering the reactor antineutrino energy range. According to Eq.~\eqref{Eq:e+NonlDetailed} and Eq.~\eqref{Eq:e+ResDetailed}, the $e^+$ energy response can be derived from the best-fit $e^-$ energy model and the measured data of $^{68}$Ge calibration source. The predictions of the $e^+$ nonlinearity and resolution, as well as the uncertainties, are shown in figure~\ref{Fig:e+Results}. 
To calculate the uncertainty bands, fitting parameters are sampled according to figure.~\ref{Fig:CorrMat} considering the full correlations for $10^4$ times. The variations of energy nonlinearity and resolution at each energy point can be evaluated from the sampled parameter sets. The uncertainty bands are defined as the symmetrical 68.3\% confidence interval of the distributions. The predictions have excellent agreements with the simulation data, where the relative residual bias for $e^+$ nonlinearity and resolution is less than $0.1\%$ and $2\%$, respectively. As a comparison of $e^+$ nonlinearity, ref.~\cite{JUNO:2020xtj} gives consistent results with residual bias within $0.2\%$ and relatively larger uncertainties.

\section{Discussions}\label{Sec:Dis}

\subsection{Impacts of fitting data inputs} \label{SubSec:inputs}
The fitting inputs for our model are $\gamma$ calibration sources and cosmogenic $^\mathrm{12}\mathrm{B}$. During the model tuning, we have found that the spectra of $\gamma$ sources are critical to constrain the resolution model due to the mono-energetic nature. However, 
the bias and uncertainty increase in the high energy region due to lack of calibration data.
Unlike the case in ref.~\cite{DayaBay:2019fje}, our Monte Carlo data does not include electronics nonlinearity, thus
the $^\mathrm{12}\mathrm{B}$ spectrum provides limited improvements on the nonlinearity fitting. Moreover, the $^\mathrm{12}\mathrm{B}$ spectrum is insensitive to the energy resolution either because of the continuous spectrum property.

\begin{figure}[!htb]
	\centering
	\includegraphics[width=0.95\hsize]{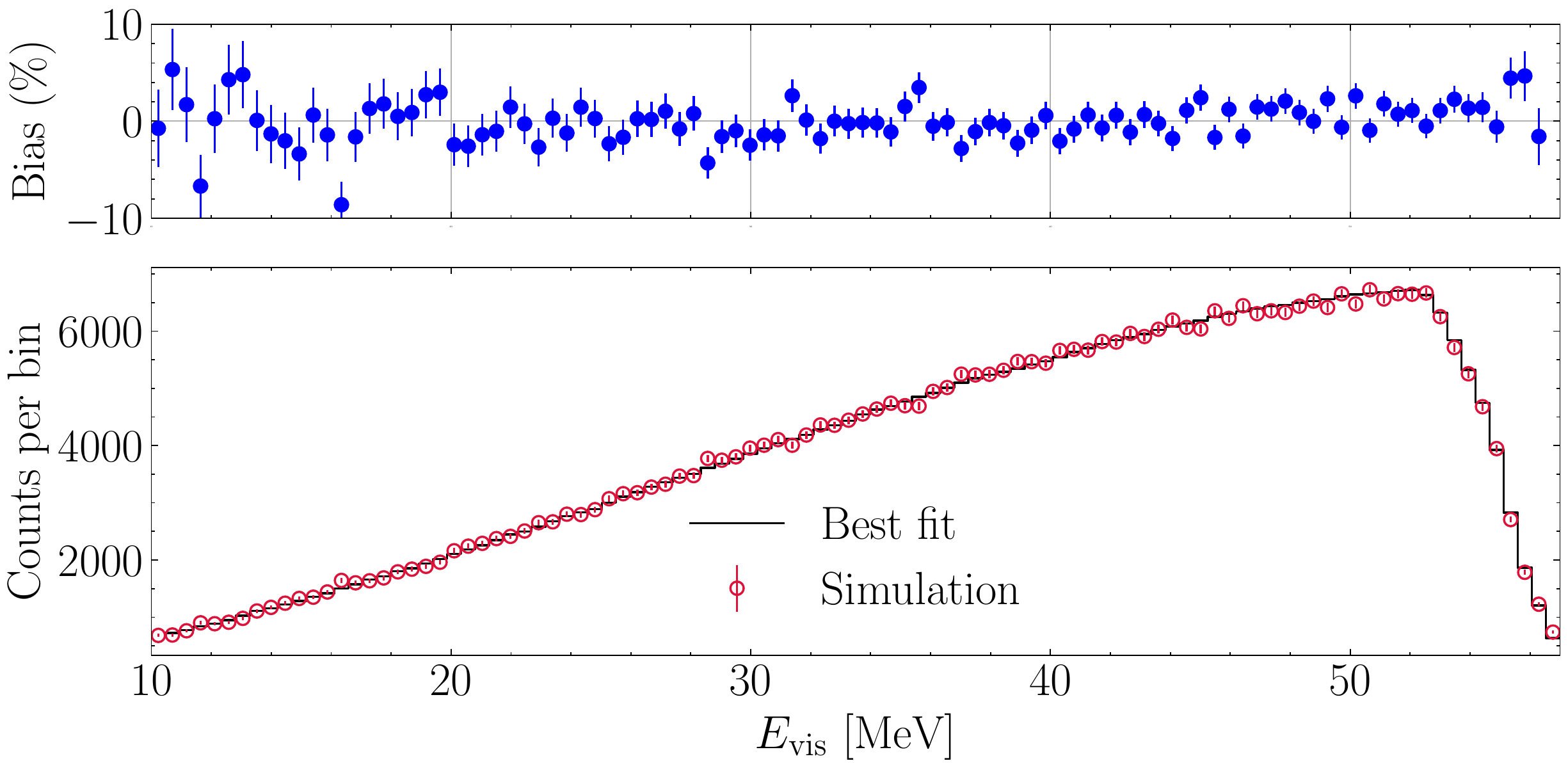}
	\caption{
	Comparison of Michel $e^-$ visible energy spectrum between the model prediction and the simulated data. The relative residual bias, with the same definition in figure~\ref{Fig:fitRes}, is shown in the upper panel.
	}
	\label{Fig:Michel}
\end{figure}

The fitting performance in the higher energy region is checked by the Michel $e^-$ samples.
The caveat is that the theoretical spectrum of Michel $e^-$ could be significantly affected if muons decay in atomic-bound states~\cite{Czarnecki:2014cxa}. Moreover, there are both $\mu^-$ and $\mu^+$ components in the cosmic muons, so the Michel $e^-$ spectrum might be contaminated by $e^+$, which introduces additional uncertainties into the model prediction. Therefore, the Michel $e^-$ spectrum is not used as a fitting input. As a test, we use the ideal energy spectrum of Michel $e^-$ without considering theoretical uncertainties or the $e^+$ contamination, and compare it with the model prediction using best-fit values in table.~\ref{Tab:FitParam}. The best-fit spectrum has good agreement with the simulated data, as shown in figure~\ref{Fig:Michel}.

In the energy region of reactor antineutrinos, our model enables a high-precision calibration of the energy nonlinearity and resolution of $e^+$. While extending to higher energy, the energy resolution is less constrained due to the lack of mono-energetic sources. This also motivates the development of new calibration strategies in the specific energy region of interest.

\subsection{Energy response separation among different particles}\label{SubSec:ParDep}
\begin{figure*}[!htb]
	\centering
	\subfigure{
    \label{Fig:AllParNonl}
    \includegraphics[width=0.46\textwidth]{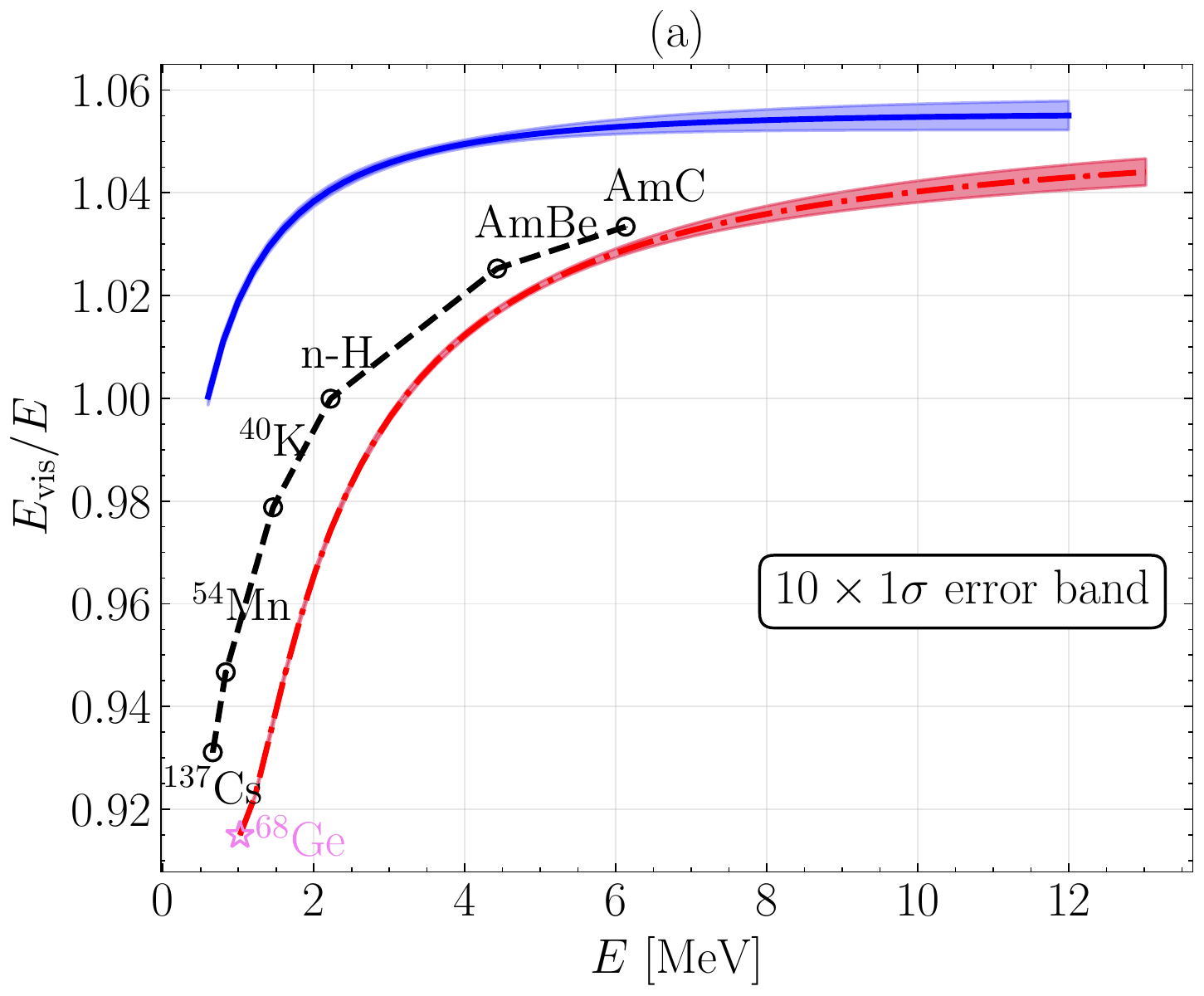}
    }
    \subfigure{
    \label{Fig:AllParResol}
    \includegraphics[width=0.46\textwidth]{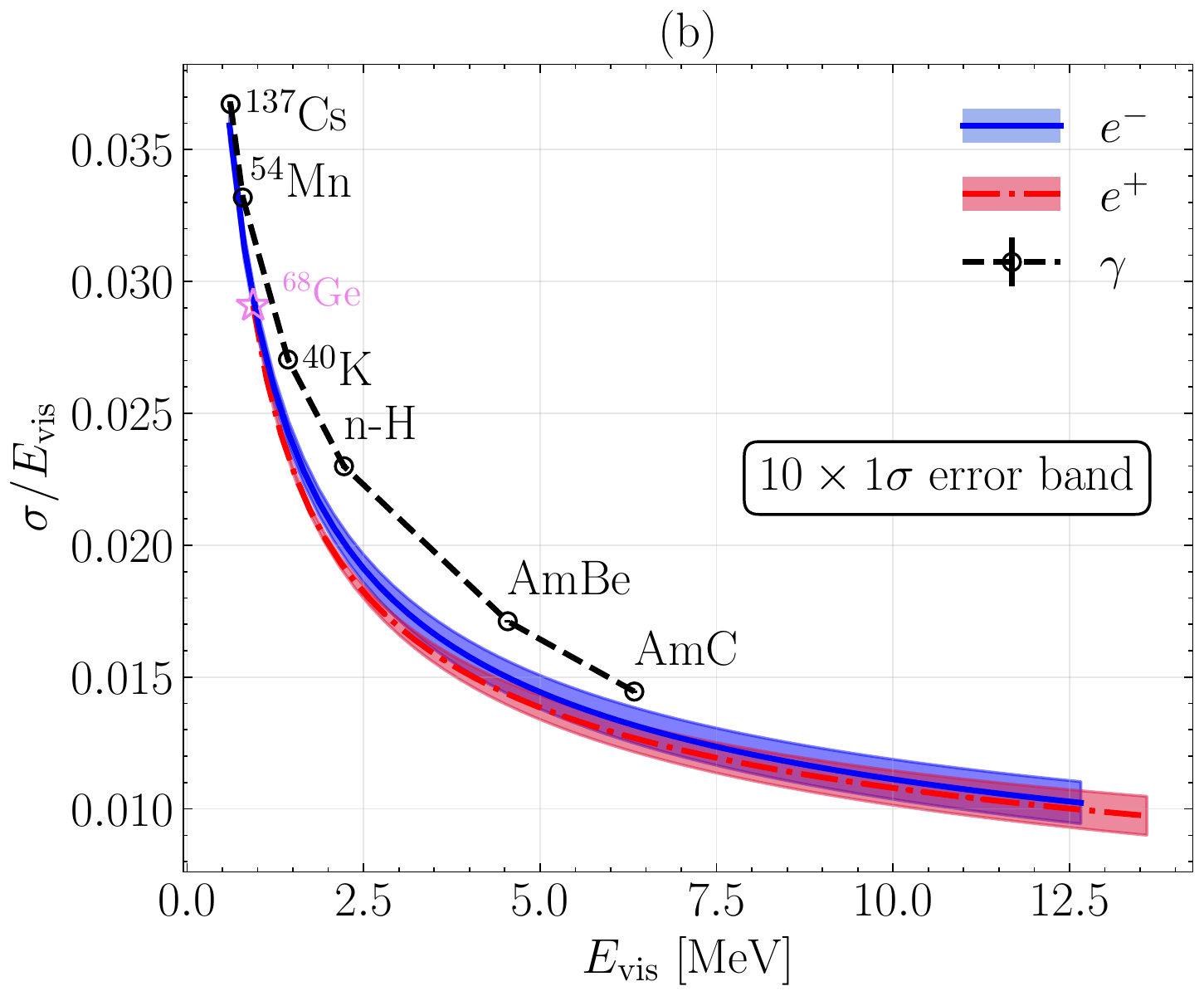}
    }
	\caption{The best-fit nonlinearity (a) and resolution (b) for $\gamma$ and $e^-$, as well as the predicted $e^+$ responses. Shadow regions on the lines represent the error bands, which have been scaled for better visualization. }
	\label{Fig:AllPar}
\end{figure*}

In figure~\ref{Fig:AllPar}, we compare the best-fit nonlinearity and resolution for $\gamma$ and $e^-$, as well as the predicted $e^+$ responses. The $^{68}$Ge calibration source anchors the starting points of the $e^+$ nonlinearity and resolution. Some observations can be made below.
\begin{itemize}
\item At the same $E_\mathrm{vis}$, the required energy of $e^-$ is lower than that of $e^+$, whereas the energy resolution of $e^-$ is slightly worse than that of $e^+$, due to more Cherenkov photons and larger non-Poisson fluctuations for $e^-$, as discussed in  section~\ref{SubSubSec:e-Resol}. 
\item The nonlinearity curve of $\gamma$ is below that of $e^-$ because its energy deposition is via multiple secondary $e^\pm$ with smaller energies. At low energies, the deposited energy of $e^+$ is dominated by the annihilation part, so $e^+$ generates fewer photons than the single $\gamma$s with the same total energy. As energy increases, the nonlinearity curve of $e^+$ will excess that of $\gamma$s.
\item The $\gamma$ resolution is notably worse than those of $e^-$ and $e^+$. It mainly results from the diversity of energy deposition modes as discussed in section~\ref{Sec:connections}. 
\item Similar to previous work~\cite{DayaBay:2019fje,JUNO:2020xtj}, the current model is able to separate the nonlinearity curves for $\gamma$ and $e^\pm$. Moreover, the new feature is the ability to obtain particle-dependent energy resolution curves, see figure~\ref{Fig:AllParResol}. High-resolution detectors like JUNO are capable of distinguishing the energy resolution between $\gamma$ and $e^+$ at the reactor antineutrino energy range.

\end{itemize}

\subsection{Model results on detectors with varied resolutions}
\begin{figure}[!htb]
	\centering
	\includegraphics[width=0.95\hsize]{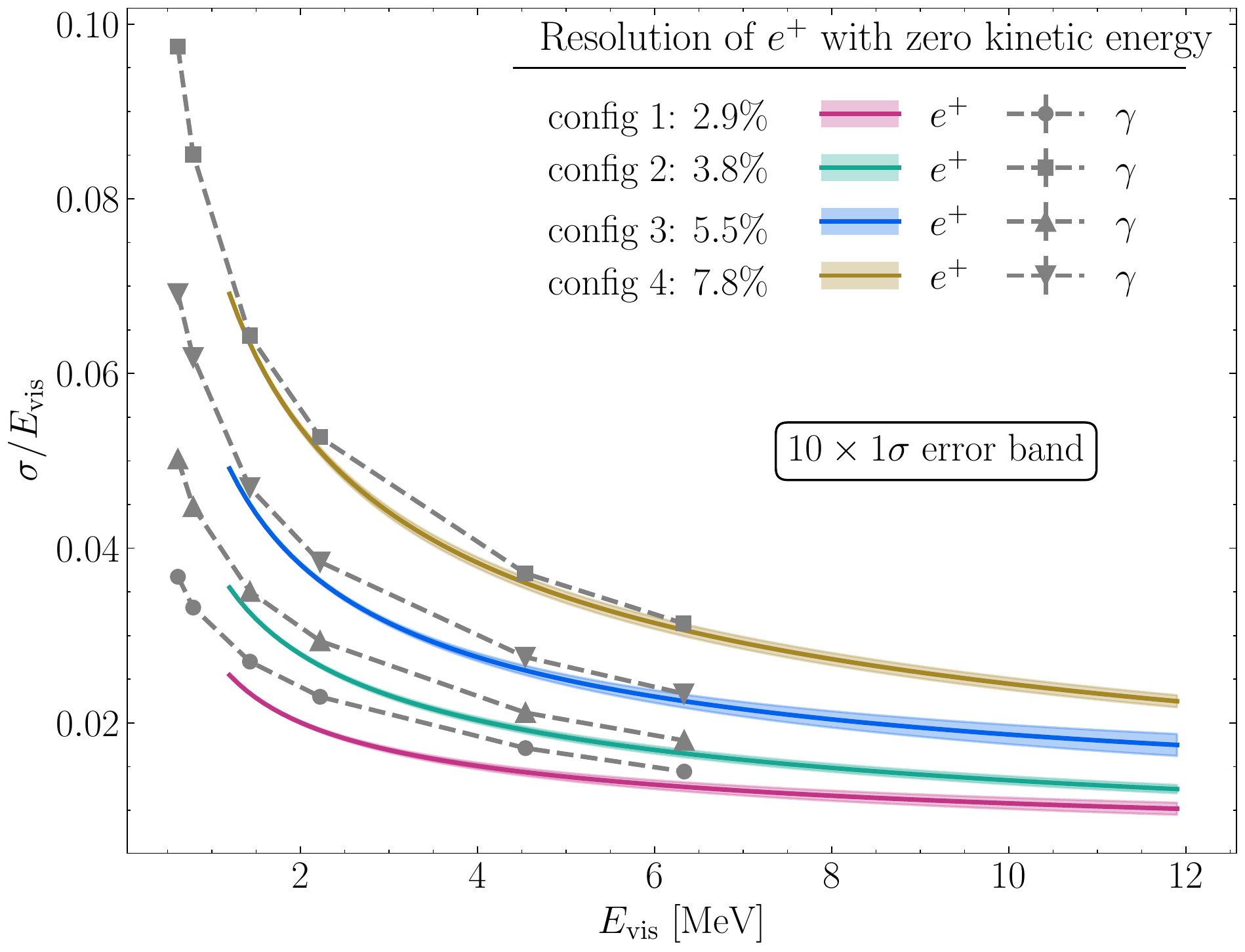}
	\caption{Fitting results of energy resolution for $e^+$ are displayed, and the shadow error regions on the lines have been scaled for better visualization. Calibration data of $\gamma$ sources are displayed together for comparison. Four different detector configurations have been studied. As the energy resolution of detectors becomes worse, the separation ability between $e^+$ and $\gamma$ decreases.}
	\label{Fig:CompareDets}
\end{figure}

By tuning the detection efficiency of the photosensors in the Monte Carlo software, it is convenient to study the model performances versus different resolutions. Four cases are simulated with the resolution values of zero kinetic energy positron being 
$2.9\%, 3.8\%, 5.5\%$ and $7.8\%$, respectively. The same fitting procedure is applied and the results are shown in figure~\ref{Fig:CompareDets}. The best-fit curves of $e^+$ resolution are shown as colored lines with error bars scaled for better visualization. For comparison, the resolution values of single $\gamma$ sources are also superimposed. The statistics of $N_\mathrm{pe}$ dominates the energy resolution variation among the four cases. The discrimination ability between $e^+$ and $\gamma$ decreases as the energy resolution of detectors becomes worse. This can be simply understood that as the LS-induced nonlinearities among all detectors are similar, the fraction of $\sigma^2_\mathrm{nonl}$ in $\left[\sigma^\gamma \right]^2$ becomes smaller for detectors with worse energy resolution according to Eq.~\eqref{Eq:gamResDecomp}. Results in figure~\ref{Fig:CompareDets} indicate that the gamma-based calibration resolution curve is incapable to approximate $e^+$ for high-resolution detectors. The proposed data-driven calibration strategy for $e^+$ resolution will strongly advance the precision measurements of reactor antineutrinos.

\section{Summary and Prospect} \label{Sec:Summary}
A comprehensive study has been carried out to construct a unified energy model for $\gamma$ and $e^\pm$ in LS detectors. 
Based on Monte Carlo studies, we have demonstrated a promising data-driven calibration approach to predict the $e^+$ nonlinearity and resolution simultaneously in the reactor antineutrino energy region. Without considering uncertainties from enclosures of $\gamma$ sources, backgrounds in the $^{12}\mathrm{B}$ spectrum and other instrumental induced effects, the relative residual bias in nonlinearity and resolution can achieve within $0.1\%$ and $2\%$, respectively. 
For higher energy region, the good agreement with simplified Michel $e^-$ spectrum has preliminarily validated the model performance, however, further exploration of dedicated calibration strategies will be needed. 
The study of energy resolution decomposition will also motivate our future developments on scintillation/Cherenkov photon discrimination algorithms.

In practice, the observed $N_\mathrm{pe}$ is often influenced by the PMT and electronics response, such as charge smearing and dark noises, and the geometric effects in LS detectors will influence the energy response such as the non-uniformity. Those instrumentally induced effects 
can be effectively eliminated by the reconstruction algorithms, e.g., see discussions in ref.~\cite{JUNO:2020xtj,Wu:2018zwk,Huang:2021baf} for JUNO. 
For the application to a realistic LS detector, above effects that impact energy response should be considered carefully and added to our model.

The JUNO-TAO experiment~\cite{JUNO:2020ijm}, which is a ton-scale satellite experiment of JUNO, aims to precisely measure the reactor antineutrino spectrum. Its targeted effective energy resolution is $<$2\% and a similar calibration strategy has been developed~\cite{Xu:2022mdi} to control the nonlinear energy scale to be $<$1\%. Our energy model would be also applicable to JUNO-TAO, and its high-resolution data could be of great value to disentangle the Cherenkov and quenching effects in energy nonlinearity and resolution.





\begin{acknowledgement}
The authors would like to thank Zeyuan Yu, Hangkun Xu and Yaoguang Wang for the helpful discussions. This work was supported in part by the National Natural
Science Foundation of China under Grant No.12125506, by the Tencent Foundation, by the National Key R\&D Program of China under Grant No.~2018YFA0404101, by the National Recruitment Program for Young Professionals, China, by the Strategic Priority Research Program of the Chinese Academy of Sciences under Grant No.~XDA10010100 and by Hubei Nuclear Solid Physics Key Laboratory.
\end{acknowledgement}




\end{document}